\documentclass[12pt,a4paper]{article}
\usepackage[textwidth=15cm]{geometry}
\usepackage{authblk}
\usepackage{amsmath} 
\usepackage{bm}        % for math
\usepackage{amssymb}   % for math
\usepackage[small]{caption}
\usepackage{epsf,epsfig,graphicx,color,hyperref}

\newcommand{\vevr}{\left< r \right>}
\newcommand{\vevh}{\left< \varphi \right>}

\newcommand{\vM}{v_{\rm M}}
\newcommand{\al}{\alpha}

\newcommand{\vk}{\vec{k}}

\newcommand{\vl}{\vec{l}}

\newcommand{\tilom}{\tilde{\omega}}

\begin{document}

\title{Effects of Goldstone Bosons on Gamma-Ray Bursts}

\author{Huitzu Tu}
%\email{huitzu2@gate.sinica.edu.tw}
%\affiliation{Institute of Physics, Academia Sinica, Taipei, Taiwan 11529, R.O.C.}
\author{Kin-Wang Ng}
%\email{nkw@phys.sinica.edu.tw}
\affil{Institute of Physics, Academia Sinica, Taipei, Taiwan 11529, R.O.C.}

%\emailAdd{huitzu@phys.sinica.edu.tw}
%\emailAdd{nkw@phys.sinica.edu.tw}

\maketitle

\begin{abstract}

Gamma-ray bursts (GRBs) are the most energetic explosion events in the universe. 
An amount of gravitational energy of the order of the rest-mass energy of the Sun is 
released from a small region, within seconds or longer.
This should lead to the formation of a fireball of temperature in the MeV range, 
consisting of electrons/positrons, photons, and a small fraction of baryons.
We exploit the potential of GRB fireballs for being a laboratory for testing particle
physics beyond the Standard Model, where we find that Weinberg's Higgs portal model
serves as a good candidate for this purpose.
Due to the resonance effects, the Goldstone bosons can be rapidly produced by 
electron-positron annihilation process in the initial fireballs of the gamma-ray bursts. 
On the other hand, the mean free path of the Goldstone bosons is larger than the size of 
the GRB initial fireballs, so they are not coupled to the GRB's relativistic flow and 
can lead to significant energy loss.
Using generic values for the GRB initial fireball energy, temperature, radius, 
expansion rate, and baryon number density, we find that the GRB bounds on 
the parameters of Weinberg's Higgs portal model are indeed competitive to current
laboratory constraints.

\end{abstract}

\maketitle

\section{Introduction}
\label{sec:intro}

Gamma-ray bursts (GRBs) are the most energetic explosion events in the 
universe (for recent reviews, see Refs.~\cite{Nakar:2007yr,Zhang:2011rp,Gehrels:2013xd,
Meszaros:2014pca,Kumar:2014upa,Pe'er:2015rfa,Meszaros:2015zka}.)
They emit a huge amount of energy of the order of 
$10^{52}~{\rm erg}$ or higher~\cite{Kumar:1999cv,Freedman:1999mp,Abdo:2009zza,
Abdo:2009pg,Wygoda:2015zua}, within a short timescale.
The initial burst of gamma-ray radiation is usually followed by an ``afterglow" at 
longer wavelengths, ranging from X-ray, optical to radio.
First detected by the military Vela satellites in late 1960's~\cite{Klebesadel:1973iq}, 
their subsequent observations by the BATSE~\cite{BATSE}, BeppoSAX~\cite{BeppoSAX}, 
Konus/Wind~\cite{Konus}, HETE-2~\cite{HETE-2}, 
Swift~\cite{Swift}, INTEGRAL~\cite{INTEGRAL}, AGILE~\cite{AGILE}, and 
Fermi~\cite{Fermi} satellites help to shed light on their nature and physical properties.
It is established that they are of cosmological origin, with the highest redshift 
recorded so far being $z = 9.4$~\cite{Cucchiara:2011pj}.
Following the investigations of Ref.~\cite{Kouveliotou:1993yx},
GRBs are commonly classified in two classes according to their $T_{90}$, the time during
which $90\%$ of the burst's fluence is accumulated.
Long bursts ($T_{90} > 2~{\rm s}$) may be due to the collapse of  
massive stars~\cite{Woosley:2006fn}, while
short bursts ($T_{90} < 2~{\rm s}$) are speculated to originate from the binary 
neutron star or neutron star - black hole mergers~\cite{Nakar:2007yr,Berger:2013jza}.
There are also ultra-long bursts~\cite{Levan:2013gcz},
or bursts whose detection requires a new classification 
scheme~\cite{Gehrels:2006tk}.

The fireball model~\cite{Cavallo:1978zz,Paczynski:1986px,Goodman:1986az,Shemi:1990rv} 
is the simplest and most conventional model to explain the observed non-thermal 
high-energy prompt emission, the variability over short timescales, and the generation 
of the afterglow of GRBs (see Refs.~\cite{Meszaros:2014pca,Kumar:2014upa,Piran:1999kx,
Meszaros:2006rc,Meszaros:2012ye,Veres:2013owa} for detailed reviews.)
In this model, the central engine is a black hole or a neutron star, surrounded by a 
matter accretion disc, which causes 
a jet of material blasted outward at relativistic speed. During the course of 
the fireball expansion, the thermal energy contained in the electrons, positrons 
and photons are gradually converted into kinetic energy of the baryons, which are
accelerated to a high Lorentz factor.
The kinetic energy is converted to gamma-ray photons in the collisions between internal 
shock waves travelling at different speeds.
At some large distances away from the central engine where the fireball becomes optically
thin, the gamma-ray photons can escape and be observed as the prompt emission.
As the shock waves continue to propagate outward, they eventually interact with the 
interstellar medium, causing the latter to emit radiations.
The long duration and the wide electromagnetic spectrum covered by those radiation 
processes then account for the observed afterglows.

The tremendous amount of energy release and the high initial temperature of the
GRB fireball makes it an excellent laboratory for particle physics.
In the Standard Model (SM), Refs.~\cite{Kumar:1999cv,Koers:2005ya} have studied the 
effects of neutrinos on the GRB initial fireballs.
It is found that although neutrino production therein is rapid enough to cool the 
fireball, the high opacity of the latter to the neutrinos 
efficiently prevents dramatic energy losses of itself~~\cite{Koers:2005ya}.
In Ref.~\cite{Derishev:1999a}, effects of the neutron component were studied in 
dependence of the final Lorentz factor of the GRB plasma wind.
It showed that neutrons can strongly influence a GRB by changing the dynamics of its
shocks in the surrounding medium. 
Beyond the SM, the possibility of using axions and other exotic particles for 
transferring the gravitational energy of the central collapsing object into the GRB 
fireball was investigated in Refs.~\cite{Loeb:1993cg,Bertolami:1998ii,Berezhiani:1999qh,
Gianfagna:2004je}, and Ref.~\cite{Demir:1999vv}, respectively.

In this work we shall show that another good example is provided by Weinberg's Higgs 
portal model~\cite{Weinberg:2013kea}, which was proposed to account for the fractional 
cosmic neutrinos in the early universe.
In this model, Weinberg considered a global $U (1)$ continuous symmetry associated with 
the conservation of some quantum number, and introduced a complex scalar field to break 
it spontaneously.
The Goldstone bosons arising from the symmetry breaking would be massless or nearly
massless, and their characteristic derivative coupling would make them very 
weakly-interacting at sufficiently low temperatures.
The latter property is crucial, since the Goldstone bosons must decouple from the early
universe thermal bath at the right moment so that their temperature is a fraction of 
that of the neutrinos (see e.g. Ref.~\cite{Ng:2014iqa}.)
We have examined energy losses due to the emission of Weinberg's Goldstone bosons in a 
post-collapse supernovae core~\cite{Keung:2013mfa}, while collider phenomenology has been 
investigated in Ref.~\cite{Cheung:2013oya}.
In this work we scrutinise the production and propagation of Weinberg's Goldstone bosons
in the initial fireballs of gamma-ray bursts.
In Section~\ref{sec:fireball} we briefly summarise generic properties of 
the GRB fireball model. 
We then review Weinberg's Higgs portal model and existing laboratory constraints on it 
in Section~\ref{sec:weinberg}.
In Section~\ref{sec:emissivity} we calculate energy loss rates due to Goldstone boson 
production by electron-positron annihilation, photon scattering, and nuclear 
bremsstrahlung processes taking place in GRB initial fireballs.
Subsequently in Section \ref{sec:mfp} we estimate the mean free path of the Goldstone 
bosons, which is set by their scattering on the electrons/positrons and nucleons.
In Section~\ref{sec:hydro} we use relativistic hydrodynamics to study the 
effects of the Goldstone bosons, and confront the results with existing laboratory 
constraints.
In Section~\ref{sec:summary} we summarise.

\section{The GRB Fireball Model}
\label{sec:fireball}

\subsection{The Fireball Mechanism}

From the correlation of the GRB duration with the progenitor environment,
it is believed that the long duration GRBs result from a collapsar, and
the short GRBs from merger.
In either case, relativistic outflows are powered by the central black hole or 
neutron star, which is surrounded by an accretion disc formed by the 
inwardly spiraling instellar material.
Two most discussed jet production mechanisms are electromagnetic extraction of 
the black hole rotation energy~\cite{Blandford:1977ds}, and pair 
annihilation of neutrinos that emanate from the accretion 
disc~\cite{Eichler:1989ve,Popham:1998ab,Zalamea:2010ax}
(see also Ref.~\cite{Globus:2014eaa} for a combination of both.)
The outcome is a large amount (of order of the solar rest mass) of gravitational 
energy released within a short time, from a small region, 
which leads to the formation of an $e^\pm$-$\gamma$ fireball.
A fraction of the gravitational energy is converted into neutrinos and 
gravitational waves. 
The thermal neutrinos are sensitive to the thermodynamics profiles of the 
accretion disc, while gravitational waves are sensitive to the dynamics of the 
progenitors. 
The Super-Kamiokande~\cite{Asakura:2015fua} and 
Sudbury Neutrino Observatory (SNO)~\cite{Aharmim:2013oba} experiments have 
searched for MeV-neutrinos from the long and short GRBs.
From the non-detection they have put upper limits on the GRB neutrino fluence.

A much smaller fraction goes into the fireball of temperature in the MeV range, 
which consists of $e^\pm$, photons and baryons, and may contain a 
comparable amount of magnetic field energy.
The initial photon luminosity inferred is many orders of magnitude larger 
than the Eddington limit, i.e. the radiation pressure far exceeds the 
gravitational force, so the fireball will expand.
For a steady spherically symmetrical flow with four velocity 
$u^\mu = (u^0, u^R, 0, 0)$ in the spherical coordinates 
$(t, R, \theta, \phi)$,
the equations of relativistic fluid dynamics are~\cite{Derishev:1999a}
\begin{equation}
   \frac{p + \rho}{n_B}\, u^0 = {\rm const.}\, , \hspace{0.4cm}
   \frac{1}{R^2} \frac{\partial}{\partial R} \left(R^2 n_B\, u^R \right) = 0\, .
\end{equation}
Here $n_B$ is the baryon number density, and $p$ and $\rho$ are the pressure 
and the total energy density, respectively. All the three quantities are 
measured in the fluid comoving frame.
The components of the flow four-velocity are $u^0 = \sqrt{ - g_{0 0}} \Gamma$
and $u^R = \beta \Gamma / \sqrt{g_{R R}}$, with $\beta$ and $\Gamma$ its
three-velocity an Lorentz factor, $\Gamma = 1 / \sqrt{1 - \beta^2}$.
If the gravitational effects of the wind itself are negligible, the metric
is $- g_{0 0} = g^{-1}_{R R} = 1 - R_{\rm S} / R$, where $R_{\rm S}$ is the
Schwarzschild radius of the central object.
The hydrodynamic equations need to be supplemented with an equation
of state, e.g. $p = \rho / 3$.
As long as the constituents of the fireball plasma are strongly coupled, they
are in thermal equilibrium, and the fireball expansion is adiabatic.
Combining with the equation of adiabatic process $p n^{- \gamma}_B = 
{\rm const.}$,
with $\gamma =4/3$, one arrives at the equation for the evolution of the 
Lorentz factor of the wind:
\begin{equation}
  \Gamma \propto R \hspace{0.3cm} {\rm for}\, \, R \ll R_{\rm sat}\, ,
  \hspace{0.6cm} \Gamma \simeq \Gamma_l \hspace{0.3cm} {\rm for}\, \,
  R \gg R_{\rm sat}\, .
\end{equation}
Here $R_{\rm sat}$ and $\Gamma_l$ are the saturation values 
for the fireball radius and the fireball Lorentz factor, respectively.
If magnetic fields are included as an additional component of the GRB fireball, 
the Lorentz factor evolution is modified to
$\Gamma (R) \propto R^\mu$ for $R < R_{\rm sat}$, and 
$\Gamma \simeq {\rm const.}$ for $R_{\rm sat} < R < R_{\rm dec}$.
Here $1/3 \leq \mu \leq 1$, with $\mu = 1$ corresponding to the baryon dominated 
jet, and $\mu = 1/3$ to a magnetic field dominated 
jet~\cite{Veres:2013owa,Veres:2012rj}. 

In any case, the Lorentz factor first increases with the radius $R$.
When $R$ reaches $R_{\rm sat} \sim 10^9~{\rm cm}$, the fireball enters 
the coasting phase, with all the fireball thermal energy converted into the 
kinetic energy of the baryons.
The fireball continues expanding at a constant rate
until it runs into the external medium and slows down.
At the deceleration radius $R_{\rm dec} \sim 10^{16}~{\rm cm}$,
the deceleration of the fireball expansion becomes significant.
Correspondingly, the fireball comoving temperature evolves as 
$T^\prime \propto R^{- \left(\frac{\mu+2}{3} \right)}$ for $R < R_{\rm sat}$, 
and $\propto R^{-2/3}$ when $R_{\rm sat} < R < R_{\rm dec}$.

The bulk Lorentz factor $\Gamma$ can be 
measured~\cite{Hascoet:2013bma,Tang:2014awa,Pe'er:2015ida}, 
and lower limits on the Lorentz factor have been inferred by requiring that
the GRBs be optically thin to high energy photons~\cite{Lithwick:2000kh}.
Ref.~\cite{Abdo:2009zza} deduced $\Gamma_{\rm min} = 608 \pm 15$ and $887 \pm 21$ 
for GRB 080916C, while $\Gamma_{\rm min} \simeq 1200$ for the short gamma-ray 
burst GRB 090510.
The saturation value for the Lorentz factor is determined by 
the initial raito of radiation energy to rest mass 
\begin{equation}
\label{eq:baryonratio}
   \eta_B \equiv \frac{\mathcal{E}}{M_0}\, .
\end{equation}
This ratio must be of the order $\sim \mathcal{O} (10^2)$, so that the 
baryons may be accelerated to a Lorentz factor 
$\Gamma \approx \mathcal{E} / M_0$ high enough to produce the observed 
gamma-rays.
On the other hand, if the ratio is too large, the fireball is radiation-dominated.
Depending on its value, there are four types of fireballs.
We consider the most interesting case, the relativistic baryonic fireball, which 
corresponds to the case $1 < \eta_B < \left(3 \sigma_T \mathcal{E} / 
8 \pi m_p R^2_0 \right)^{1/3} \approx 10^5\, 
\left(\mathcal{E} / 10^{52}~{\rm ergs} \right)^{1/3}\, 
\left(R_0 / 10^7~{\rm cm} \right)^{-2/3}$~\cite{Shemi:1990rv,Piran:1999kx},
where $\sigma_T$ is the Thompson cross section, and $m_p$ the proton mass.

Within the fireball model, there are many mechanisms proposed to explain the 
GRB observations. 
In the internal-external scenario, the prompt emission is produced by
the internal shocks~\cite{Rees:1994nw}, and the afterglow by the external shocks.
Under the assumption that the central engine produces ejecta shells with a 
highly variational distribution of Lorentz factor, the internal shocks are 
formed when the faster shells catch up with the slower ones.  
The external shocks arises when the fireball expands into external medium.
For a summary or review of the GRB fireball model, we refer to 
Refs.~\cite{Meszaros:2014pca,Kumar:2014upa,Piran:1999kx,Meszaros:2006rc,Meszaros:2012ye,
Veres:2013owa}.

\subsection{Generic GRB Fireball Parameters}

In this work we consider the following generic parameters for the GRB fireballs
as Ref.~\cite{Koers:2005ya}:
the initial fireball energy is ${\cal E} = 10^{52}$ - $10^{54}~{\rm ergs}$.
The initial radius is that of the Schwarzschild radius 
$R_{\rm S} = 3 \left(M / M_\odot \right)~{\rm km}$, or of the  
neutron star radius $\sim 10~{\rm km}$.
The initial wind velocity is about the sound speed, 
$\beta_0 \approx c_s = 1 /\sqrt{3}$.
In thermal equilibrium, the radiation energy density and the temperature is 
related by
\begin{equation}
   \frac{\mathcal{E}}{V} = \frac{\pi^2}{30}\, g_\ast T^4\, .
\end{equation}
The total number of effective massless degrees of freedom is 
\begin{equation}
   g_\ast \equiv \sum_{i = {\rm bosons}} g_i \left(\frac{T_i}{T} \right)^4
   + \frac{7}{8} \sum_{i = {\rm fermions}} g_i \left(\frac{T_i}{T} \right)^4\, ,
\end{equation}
with $g_i$ the internal degrees of freedom of particle species $i$, and
$T_i$ its temperature.
In the initial fireball, photons, electrons, positrons, as well as
three flavours of neutrinos are in thermal equilibrum, so $g_\ast = 43/4$.
Assuming that the initial fireball is spherical,
its temperature can be expressed by
\begin{equation}
\label{eq:T0_spherical}
   \left(T_{11} \right)^4 = \frac{200}{g_\ast} \frac{\mathcal{E}_{52}}
   {\left(R_{\rm 6.5} \right)^3}\, ,
\end{equation}
where $T = T_{11} \times 10^{11}~{\rm K}$, 
$\mathcal{E} = \mathcal{E}_{52} \times 10^{52}~{\rm erg}$, and
$R = R_{6.5} \times 10^{6.5}~{\rm cm}$.
We therefore follow Ref.~\cite{Koers:2005ya} to choose 
\begin{equation}
   \mathcal{E} = 10^{52}~{\rm erg}\, , \hspace{0.4cm}
   R_0 = 10^{6.5}~{\rm cm}\, , \hspace{0.4cm}
   T_0 = 2.1 \cdot 10^{11}~{\rm K} = 18~{\rm MeV}\, ,
\end{equation}
as our fiducial value for the initial fireball total energy, radius and 
temperature, respectively.
In view of the recent results in Ref.~\cite{Pe'er:2015ida}, we also consider larger
initial radius, e.g. $R_0 = 10^7$ - $10^8~{\rm cm}$, and lower initial fireball 
temperature, values, such as $T_0 = 8$ and $2~{\rm MeV}$.

It was shown (see e.g. Ref.~\cite{Levinson:2013mwa})
that the sonic point of a Schwarzschild black hole should be located at the 
radius $R_c = \frac{3}{2} R_{\rm S}$, if the particles in the in- and 
outflow are relativistic so that the equation of state is $p = \rho /3$.
In the case that the GRB jets are formed by energy injection from neutrino
pair annihilation, the sonic point of the inflow,
$R_{c, 1} < \frac{3}{2} R_{\rm S}$ (where $\beta_{c, 1} = - \frac{1}{\sqrt{3}}$), 
and that of the outflow, $R_{c, 2} > \frac{3}{2} R_{\rm S}$
(where $\beta_{c, 2} = \frac{1}{\sqrt{3}}$), are separate. 
In Ref.~\cite{Levinson:2013mwa} the location of the outer sonic point is
shown for several different energy injection profiles, which is pushed 
out well above that of the adiabatic flow ($R_c = \frac{3}{2} R_{\rm S}$) 
in all cases (see also Ref.~\cite{Derishev:1999a}.)
In this work we choose 
\begin{equation}
   \beta_0 \left( R = R_0 \right) = c_s = \frac{1}{\sqrt{3}}\, , 
\end{equation}
as the fiducial value for the fireball initial wind velocity.

Since the initial temperature is higher than the nuclear binding energies,
the nuclei are dissociated in nucleons.
Requiring $\eta_B \sim 1000$ for the initial energy to rest mass ratio
defined in Eq.~(\ref{eq:baryonratio}), the 
initial comoving baryon number density in the fireballs should be
\begin{equation}
   n_{B, 0} = 5 \cdot 10^{31}~{\rm cm}^{-3}\, , 
\end{equation}
so that the fireball rest mass 
$M_0 = m_N n_{B, 0} V_0 \approx 10^{49}~{\rm ergs}$.
The electron and the positrion number density are
\begin{equation}
\label{eq:fepm}
   n_{e^\pm} = 2 \int f_{e^\pm} (\vec{p}_{e^\pm})\, 
   \frac{d^3 \vec{p}_{e^\pm}}{(2 \pi)^3}\, ,
\end{equation}
with their phase space distribution functions given by
$f_{e^-} (\vec{p}_e^-) = \left(e^{\left( E_{e^-} - \mu_e \right) / T} + 
1 \right)^{-1}$ and 
$f_{e^+} (\vec{p}_e^+) = \left(e^{\left( E_{e^+} + \mu_e \right) / T} + 
1 \right)^{-1}$, respectively.
The $e^\pm$ chemical potential $\mu_e$ is determined by the requirement of
charge neutrality and beta-equilibrium in the fireball for a fixed lepton
fraction $Y_e$. 
For the reference temperature, it is 
$\mu_e / T_0 \sim 2 \times 10^{-4}$~\cite{Koers:2005ya}, i.e. the electrons
and positrons are non-degenerate, so
\begin{equation}
   n_{e^\pm, 0} = \frac{3}{4} \frac{\zeta (3)}{\pi^2}\, 2 T^3 
   = 1.4 \cdot 10^{35} \left( \frac{T}{18~{\rm MeV}} \right)^3~{\rm cm^{-3}}\, ,
\end{equation}
with $\zeta (3) \approx 1.20206$.
Neutrinos are created rapidly in the initial fireball, majorly through
the electron-positron pair annihilation process 
$e^- + e^+ \rightarrow \nu + \bar{\nu}$.
The emissivity for this process is~\cite{Koers:2005ya,Itoh:1989a}
\begin{equation}
   Q_{\rm e^- e^+ \rightarrow \nu_i \bar{\nu}_i} = 3.6 \cdot 10^{33}\,
   \left(T_{11} \right)^9~{\rm erg}~{\rm s}^{-1}~{\rm cm}^{-3}\, ,
\end{equation}
much larger than that for the photo-neutrino
$e^\pm + \gamma \rightarrow e^\pm + \nu_i + \bar{\nu}_i$, the plasma
$\gamma \rightarrow \nu_i \bar{\nu}_i$, and the URCA processes
$e^- + p \rightarrow n + \nu_e$ and $e^+ + n \rightarrow p + \bar{\nu}_e$.
Neutrino mean free path (mfp) is set by the elastic scattering on electrons
and positrons $\nu + e^\pm \rightarrow \nu + e^\pm$. 
It is~\cite{Koers:2005ya}
\begin{equation}
   \lambda^{(e)} = 3.7 \cdot 10^6 \left(T_{11} \right)^{-5}~{\rm cm}\, , 
   \hspace{0.5cm}
   \lambda^{(\mu, \tau)} = 1.6 \cdot 10^7 
   \left(T_{11} \right)^{-5}~{\rm cm}\, ,
\end{equation} 
for the three flavours, respectively.
Neutrinos decouple in two stages, when the optical depth 
($\tau \equiv R / \lambda$) for each neutrino
flavour, $\tau^{(\mu, \tau)}$ and $\tau^{(e)}$, drops to 1.

In this work we consider a baryon dominated fireball jet, and neglect the
effects of the magnetic fields.

\section{Goldstone Bosons from Weinberg's Higgs Portal Model}
\label{sec:weinberg}

\subsection{The Model}

In this subsection we briefly summarise Weinberg's model~\cite{Weinberg:2013kea} 
following the convention of Refs.~\cite{Keung:2013mfa,Cheung:2013oya}.
Consider the simplest possible broken continuous symmetry, a global $U (1)$ symmetry
associated with the conservation of some quantum number $W$.
A single complex scalar field $S (x)$ is introduced for breaking this symmetry
spontaneously. 
With this field added to the Standard Model (SM), the Lagrangian is
\begin{equation}
\label{eq:Lagrangian1}
  {\mathcal L} = \left(\partial_\mu S^\dagger \right) \left(\partial^\mu S \right)
  + \mu^2 S^\dagger S - \lambda (S^\dagger S)^2 - g (S^\dagger S) 
  (\Phi^\dagger \Phi) + {\mathcal L}_{\rm SM}\, ,
\end{equation}
where $\Phi$ is the SM Higgs doublet, $\mu^2$, $g$, and $\lambda$ are real constants,
and $\mathcal{L}_{\rm SM}$ is the usual SM Lagrangian.
One separates a massless Goldstone boson field $\alpha (x)$ and a massive radial field
$r (x)$ in $S (x)$ by defining
\begin{equation}
   S (x) = \frac{1}{\sqrt{2}} \left(\vevr + r (x) \right)\, e^{2 i \al (x)}\, ,
\end{equation}
where the fields $\alpha (x)$ and $r (x)$ are real.
In the unitary gauge, one sets $\Phi^{\rm T} = \left(0, \vevh + \varphi (x) \right) 
/\sqrt{2}$ where $\varphi (x)$ is the physical Higgs field.
The Lagrangian in Eq.~(\ref{eq:Lagrangian1}) thus becomes
\begin{eqnarray}
   \mathcal{L} &=& \frac{1}{2} \left(\partial_\mu r \right) \left(\partial^\mu r \right)
   + \frac{1}{2} \frac{\left(\vevr + r \right)^2}{\vevr^2} 
   \left(\partial_\mu \al \right) \left(\partial^\mu \al \right) +
   \frac{\mu^2}{2} \left(\vevr + r \right)^2 \nonumber \\
   && - \frac{\lambda}{4} \left(\vevr + r \right)^4 - \frac{g}{4}
   \left(\vevr + r \right)^2 \left(\vevh + \varphi \right)^2 + \mathcal{L}_{\rm SM}\, ,
\end{eqnarray}
where the replacement $\al (x) \rightarrow \al (x) / \left( 2 \vevr \right)$ was made 
in order to achieve a canonical kinetic term for the $\al (x)$ field.
In this model, the interaction of the Goldstone bosons with the SM particles arises
entirely from a mixing of the radial boson with the Higgs boson.
The mixing angle is
\begin{equation}
\label{eq:mixingangle}
   \tan 2 \theta = \frac{2 g \vevh \vevr}{m^2_\varphi - m^2_r}\, .
\end{equation}
Collider searches for the SM Higgs invisible decay as well as meson invisible decays
have already been used to set strong constraints on the coupling $g$ and/or the 
mixing angle $\theta$, as will be reviewed in the next subsection.

\begin{center}
\begin{figure}[t!]
\includegraphics[width=0.6\textwidth,angle=-90]{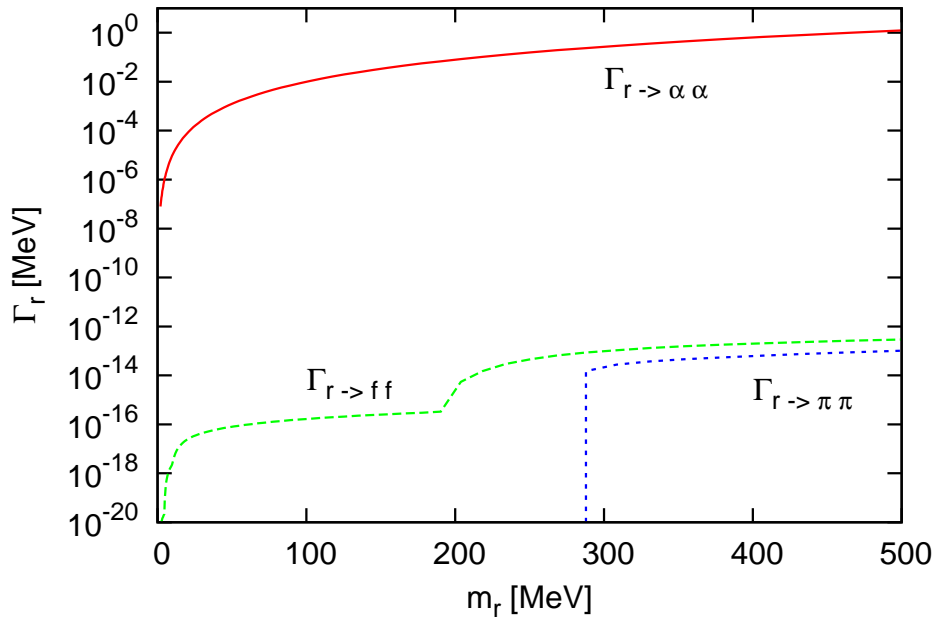}
\caption{Decay widths for the $r$ field to Goldstone boson pairs (solid), 
fermion pairs (dashed), and pion pairs (dotted), assuming $g = 0.011$ and 
$\vevr = 1~{\rm GeV}$.}
\label{fig:Gamma_r}
\end{figure}
\end{center}

As will be presented in Section~\ref{sec:emissivity},
Goldstone boson emissivities in the GRB initial fireballs depends strongly on the 
total decay width of the radial field $r$.
The $r$ field decays dominantly to a pair of Goldstone bosons, with 
the decay width given by
\begin{equation}
   \Gamma_{r \rightarrow \al \al} = \frac{1}{32 \pi}\, \frac{m^3_r}{\vevr^2}\, .
\end{equation}
However, if its vacuum expectation value $\vevr$ is very large and the 
coupling $g$ is not too small, the decay widths into SM fermion pairs
\begin{equation}
   \Gamma_{r \rightarrow f \bar{f}} = \left(\frac{g \vevh \vevr}
   {m^2_r - m^2_\varphi} \right)^2 \frac{N_c\, m^2_f\, m_r}{8 \pi \vevh^2}
   \left(1 - \frac{4 m^2_f}{m^2_r} \right)^{3/2}\, ,
\end{equation}
can be comparable or even dominant.
Here $m_f$ and $N_c$ are the mass and the colour factor of the fermion, respectively.
For $m_r > 2 m_\pi$, the $r$ field can also decay to pion pairs through the effective 
coupling of the SM Higgs to pions $\left< \pi^+ \pi^- \vert \mathcal{L}_{\rm int} 
\vert \varphi \right>$.
The effective Lagrangian is~\cite{Vainshtein:1980ea,Voloshin:1985tc,Gunion:1989we}
\begin{equation}
\label{eq:Lhiggspion}
  \mathcal{L}_{\rm int} = - \frac{\varphi}{\vevh} \Big\{ \frac{2}{9} \theta^\mu_\mu
  + \frac{7}{9} \sum_{ i = u, d, s} m_i \bar{\phi}_i  \phi_i \Big\}\, ,
\end{equation}
where $\theta^\mu_\mu$ is the trace of the energy-momentum tensor, valid only
at low momentum tranfers $\lesssim 0.3~{\rm GeV}$ (see e.g. the discussion in 
Ref.~\cite{Ng:2014iqa}.)
The decay width is then
\begin{equation}
  \Gamma_{r \rightarrow \pi^+ \pi^-} = \left( \frac{g \vevh \vevr}
  {m^2_r - m^2_\varphi} \right)^2 \left(\frac{2}{9 \vevh} \right)^2
  \frac{\left( m^2_r + \frac{11}{2} m^2_\pi \right)^2}{16 \pi m_r} 
  \left(1 - \frac{4 m^2_\pi}{m^2_r} \right)^{1/2}\, .
\end{equation}
In Fig.~\ref{fig:Gamma_r} the three decay widths are shown for the case of 
$\vevr = 1~{\rm GeV}$, $g = 0.011$.
The decay widths for other parameter values can be easily obtained by scaling with
$g^2$ or $\vevr^2$.

In the GRB fireballs, the Goldstone bosons can also be produced by nuclear processes and
undergo elastic scattering with the nucleons through the $\varphi-r$ mixing.
The Higgs effective coupling to nucleons, $f_N m_N / \vevh$, has been calculated 
for the purpose of investigating the sensitivities of the dark matter direct detection 
experiments~\cite{Drees:1993bu,Jungman:1995df,Hisano:2011cs,Cline:2013gha}.
Following the Shifman-Vainshtein-Zakharov approach~\cite{Shifman:1978zn} to evaluate 
the contributions from the heavy quarks, it can be written in the form
\begin{equation}
   f_N \equiv \sum_{q = u, d, s, c, b, t} \frac{m_q}{m_N} \left< N \vert \bar{q} 
   q \vert N \right> = \sum_{q = u, d, s} f_{T q} + \frac{2}{9} \left[1 - 
   \sum_{q=u, d, s} f_{T q} \right]\, .   
\end{equation}
In this work we use the estimate of $f_N = 0.3$ for proton and neutron from 
Ref.~\cite{Cline:2013gha}.

\subsection{Laboratory Constraints}
\label{sec:labconstraints}

\subsubsection{Collider Searches for SM Higgs Invisible Decay}

The non-standard decay branching ratio of the SM Higgs is constrained to 
$\Gamma_{h \rightarrow {\rm inv.}} < 0.97~{\rm MeV}$ at the $95\%$ confidence level
(corresponding to a branching ratio $< 19\%$) by the results of a global fitting to
the most updated data from the CMS and ATLAS experiments at the Large Hadron Collider
(LHC), as well as those from the Tevatron~\cite{Cheung:2014noa}.
In Weinberg's Higgs portal model, the SM Higgs can decay into a pair of Goldstone bosons 
or a pair of the radial field $r$, with the decay widths given by
\begin{equation}
   \Gamma_{\varphi \rightarrow \al \al} = \frac{1}{32 \pi}\, 
   \frac{g^2 \vevh^2 m^3_\varphi}{(m^2_\varphi - m^2_r)^2}\, , \hspace{0.3cm}
   {\rm and} \hspace{0.2cm}
   \Gamma_{\varphi \rightarrow r r } = \frac{ g^2 \vevh^2}{32 \pi}\,  
   \frac{\sqrt{m^2_\varphi - 4 m^2_r}}{m^2_\varphi}\, ,
\end{equation}
respectively.
The constraint is translated into a bound on the Goldstone boson coupling of
\begin{equation}
   |g| < 0.011\, . 
\end{equation}
In the future, the International Linear Collider (ILC) may reach a sensitivity of 
constraining the branching ratio of SM Higgs invisible decays to 
$< 0.4 - 0.9\%$~\cite{Bechtle:2014ewa} in the best scenarios.
If this can be realised, the collider bound on the Goldstone boson coupling will be 
improved by a factor of $5 \sim 7$.
In this work we will estimate the effects of the Goldstone bosons on the initial 
GRB fireballs for the coupling in the range $0.011 > g > 0.0015$.

\subsubsection{Muon Anomalous Magnetic Moment}

There is still a discrepancy between the SM prediction for the muon anomalous 
magnetic moment~\cite{Blum:2013xva}, $a^{\rm SM}_\mu$, and the experimental 
results from the E821 experiment at Brookhaven National Lab (BNL)~\cite{Bennett:2006fi},
$a^{\rm exp}_\mu = 11\, 659\, 209 (5.4) (3.3) \cdot 10^{-10}$, 
where the first errors are statistical and the second systematic.
The observed difference of~\cite{Agashe:2014kda}
\begin{equation}
   \Delta a_\mu = a^{\rm exp}_\mu - a^{\rm SM}_\mu = 288 (63) (49) 
   \cdot 10^{-11}\, ,
\end{equation}
may point to new physics beyond the Standard Model.
The contribution from the SM Higgs was first calculated in 
Ref.~\cite{Jackiw:1972jz}.
The radial $r$ field can contribute to $\Delta a_\mu$ through its mixing with the 
SM Higgs~\cite{Huang:2013oua}
\begin{equation}
   \Delta a^r_\mu = \theta^2\, \frac{G_F\, m^2_\mu}{4 \pi^2 \sqrt{2}}
   \int^1_0 d y \frac{y^2 (2 - y)}{y^2 + (1 - y) \left( \frac{m_r}{m_\mu} 
   \right)^2}\, ,
\end{equation}
where $G_F$ is the Fermi constant.
By demanding $\Delta a^r_\mu < \Delta a_\mu$ one obtains a very weak constraint on
the mixing angle: $\theta \lesssim \mathcal{O} (1)$.

\subsubsection{Radiative Upsilon Decays}

As first pointed out by Wilczek~\cite{Wilczek:1977pj}, light Higgs boson 
can be searched for in the radiative decays of heavy vector mesons.
In Weinberg's Higgs portal model, the branching ratio is
\begin{equation}
   \frac{\mathcal{B} \left(\Upsilon (n S) \rightarrow \gamma + r \right)}
   {\mathcal{B} \left(\Upsilon (n S) \rightarrow \mu^+ \mu^- \right)}
   = \theta^2 \frac{G_F\, m^2_b}{\sqrt{2} \pi \alpha}\, 
   \left(1 - \frac{m^2_r}{m^2_{\Upsilon (n S)}} \right) \times F\, ,
\end{equation}
for $n = 1, 2, 3$, where $\alpha$ is the fine structure constant, and 
$m_b$ the $b$ quark mass.
The correction factor $F$ includes QCD and relativistic corrections 
(see e.g. Refs.~\cite{Gunion:1989we,Huang:2013oua,Domingo:2008rr}.)
For $m_r \ll m_{\Upsilon}$, $F \sim 0.5$ is generally assumed. 
The BaBar Collaboration has set $90\%$ C.L. upper limits on 
$\mathcal{B} \left(\Upsilon (1 S) \rightarrow \gamma A^0 \right) 
\times \mathcal{B} (A^0 \rightarrow {\rm invisible})$ in the range
$(1.9 - 37) \times 10^{-6}$ for 
$m_{A^0} < 9.2~{\rm GeV}$~\cite{delAmoSanchez:2010ac}, as well as
on $\mathcal{B} \left(\Upsilon (3 S) \rightarrow \gamma A^0 \right) 
\times \mathcal{B} \left(A^0 \rightarrow {\rm invisible} \right)$ in the range
$(0.7 - 31) \times 10^{-6}$ for 
$m_{A^0} < 9.2~{\rm GeV}$~\cite{Aubert:2008as}, where $A^0$ is a scalar boson.
In this work we consider mass of the radial field $r$ below $1~{\rm GeV}$,
for which
\begin{equation}
   \mathcal{B} \left(\Upsilon (n S) \rightarrow \gamma + r \right) 
   < 3 \times 10^{-6}\, ,
\end{equation}
for $n = 1$ and 3.
This is translated into a constraint on Weinberg's Higgs portal model as
$\theta < 0.2$.

\subsubsection{$\textbf{B}$ Meson and Kaon Decays}

As first pointed out in Ref.~\cite{Bird:2004ts}, decays of $B$ mesons to a $K$ meson
plus missing energy can be an efficient probe of GeV or sub-GeV scalar dark matter.
In Refs.~\cite{Huang:2013oua,Anchordoqui:2013bfa} this constraint has been applied 
to Weinberg's Higgs portal model.
The BaBar Collaboration has reported an upper limit at the 
$90\%$ confidence level of $\mathcal{B} (B^+ \rightarrow K^+ + \nu \bar{\nu}) < 1.3 
\cdot 10^{-5}$, as well as
$\mathcal{B} (B^0 \rightarrow K^0 + \nu \bar{\nu}) 
< 5.6 \cdot 10^{-5}$~\cite{delAmoSanchez:2010bk}.
The CLEO Collaboration also reported a $90\%$ C. L. upper limit of 
$4.9 \cdot 10^{-5}$ and $5.3 \cdot 10^{-5}$ on the branching ratio for the decays
$B^\pm \rightarrow K^\pm X^0$ and $B^0 \rightarrow K^0_S X^0 $, 
respectively, where $X^0$ is any neutral massless weakly-interacing 
particle~\cite{Ammar:2001gi}.
In the SM, the branching ratio for the total 
${\mathcal B} (B \rightarrow K  \nu \bar{\nu})$ branching ratio is estimated to
be  $(4.5 \pm 0.7) \cdot 10^{-6}$.
In Weinberg's Higgs portal model, this branching ratio  
is~\cite{Huang:2013oua,Anchordoqui:2013bfa}
\begin{eqnarray}
\label{eq:Bdecay}
   \mathcal{B} (B^+ \rightarrow K^+ + r) &=& 
   \frac{9 \sqrt{2} \tau_B\, G^3_F\, m^4_t m^2_b}{1024 \pi^5 m^3_B} 
   \frac{m^2_+ m^2_-}{\left(m_b - m_s \right)^2}
   |V_{t b}\, V^\ast_{t s}|^2\, \left[f^{B K}_0 (m^2_r) \right]^2 \nonumber \\
   && \hspace{0.5cm} \times \sqrt{\left(m^2_+ - m^2_r \right) 
   \left(m^2_- - m^2_r \right)}\, \theta^2\, ,
\end{eqnarray}
with the form factor given by
\begin{equation}
   f^{B K}_0 (x) \equiv 0.33 \exp \left(\frac{0.63 x}{m^2_B} - \frac{0.095 x^2}{m^4_B} 
   + \frac{0.591 x^3}{m^6_B} \right)\, ,
\end{equation}
and $\tau_B$ is the $B$-meson lifetime, $V_{t b}$ and $V_{t s}$ the CKM matrix 
elements, and $m_t$, $m_b$, $m_s$, $m_B$ and $m_K$ the corresponding quark and meson
masses, with $m_{\pm} \equiv m_B \pm m_K$.
We follow Ref.~\cite{Anchordoqui:2013bfa} and use the most stringent constraint
\begin{equation}
   \mathcal{B} (B^+ \rightarrow K^+ + r) < 10^{-5}\, ,
\end{equation}
which imposes a constraint on the $\varphi - r$ mixing angle that $\theta < 0.0016$, 
for $m_r < m_B - m_K$.

If the radial field $r$ is lighter than $354~{\rm MeV}$, the  
decay of K meson to a $\pi$ meson plus missing energy is a more powerful probe.
The E787 and E949 experiments at the BNL has used stopped kaons to study the
rare decay $K^+ \rightarrow \pi^+ \nu \bar{\nu}$ ~\cite{Artamonov:2009sz}.
The branching ratio $\mathcal{B} (K^+ \rightarrow \pi + \nu \bar{\nu}) = \left(
1.73^{+1.15}_{-1.05} \right) \cdot 10^{-10}$ determined with the observed seven events 
and background estimation is consistent with the SM prediction of 
$7.8 (75) (29) \cdot 10^{-11}$~\cite{Buchalla:1998ba,Brod:2010hi}, where the 
first error summarises the parametric, and the second the remaining theoretical 
uncertainties.
For $K^+$ meson decay into the radial $r$ field, the branching ratio 
can be calculated similarly as in Eq.~(\ref{eq:Bdecay}), using the form factor 
(see e.g. Ref.~\cite{Anchordoqui:2013bfa}),
\begin{equation}
   f^{K \pi}_0 (x) \approx 0.96 \left(1 + 0.02 \frac{x}{m^2_\pi} \right)\, .
\end{equation}
In this work we follow Refs.~\cite{Huang:2013oua,Anchordoqui:2013bfa} and use the
constraint
\begin{equation}
   \mathcal{B} (K^+ \rightarrow \pi^+ + r) < 10^{-10}\, , 
\end{equation}
which imposes a very stringent constraint on the mixing angle as 
$\theta < 8.7 \cdot 10^{-5}$, for $m_r < m_K - m_\pi = 354~{\rm MeV}$.

Laboratory constraints from muon anomalous magnetic moment, radiative upsilon decays, 
as well as $B^+$ and $K^+$ invisible decays are plotted in 
Fig.~\ref{fig:GRBmeson_gcoupvevr}, in terms of upper limits on $g \vevr$, the product
of the Goldstone boson coupling times the vev of the $r$ field, versus its mass $m_r$
(cf. Eq.~(\ref{eq:mixingangle})).

\section{GRB Energy Losses due to Goldstone Boson Production}
\label{sec:emissivity}

The Goldstone bosons can be produced in electron-positron pair annihilation  
$e^+ + e^- \rightarrow \al + \al$, photon scattering $\gamma + \gamma \rightarrow 
\al + \al$, as well as in nuclear bremsstrahlung processes
$N + N \rightarrow N + N + \al + \al$.

\subsection{Electron-Positron Pair Annihilation}

The amplitude squared for $e^+ (p_1)\, e^- (p_2) \rightarrow \al (q_1)\, \al (q_2)$ is
\begin{equation}
   \sum_{\rm spins} \vert {\cal M}_{e^+ e^- \rightarrow \al \al}\vert^2 =
   \frac{2 g^2\, m^2_e}{m^4_\varphi}
   \frac{ s^2 (s - 4 m^2_e)}
   {\left(s - m^2_r \right)^2 + m^2_r \Gamma^2_r}\, ,
\end{equation}
where $s = \left(p_1 + p_2 \right)^2 = \left(q_1 + q_2 \right)^2$ is the centre-of-mass 
(cm) energy.
Here we have used the Breit-Wigner form to take into account the resonance effect, with
the $r$ field total decay width 
$\Gamma_r = \Gamma_{r \rightarrow \al \al} + \sum_{f | 2 m_f < m_r} 
\Gamma_{r \rightarrow f \bar{f}} + \Gamma_{r \rightarrow \pi^+ \pi^-}$ 
(cf. Fig.~\ref{fig:Gamma_r}).  

\begin{center}
\begin{figure}[t!]
\includegraphics[width=0.6\textwidth,angle=-90]{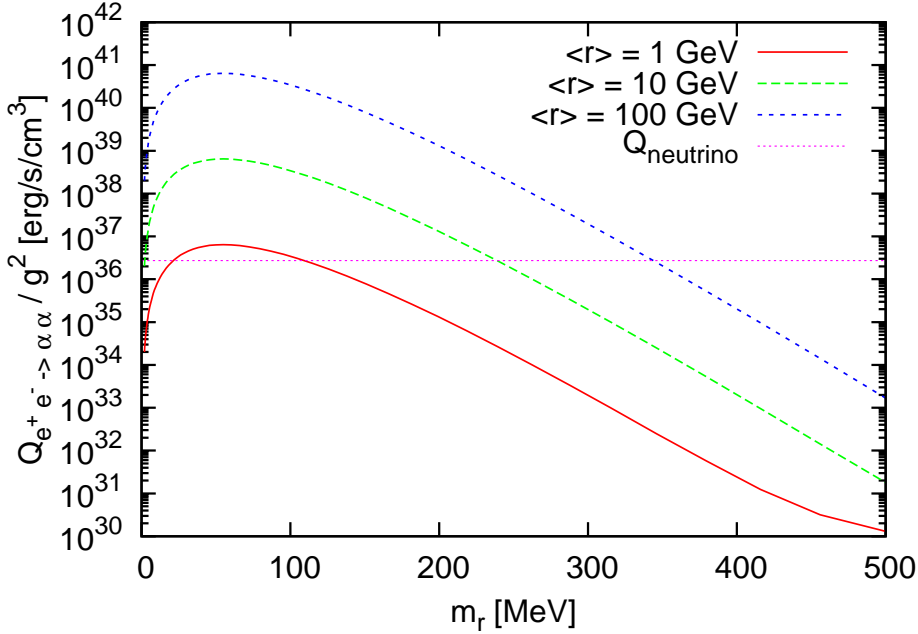}
\caption{Energy loss rate due to Goldstone boson production from 
$e^- e^+ \rightarrow \al \al$ divided by the Goldstone boson coupling $g^2$ vs.
the radial boson mass $m_r$.
The GRB initial fireball temperature is set at the fiducial value
$T_0 = 18~{\rm MeV}$, and the vacuum expectation value of the radial boson is
assumed to be $\vevr=1$, $10$, and $100~{\rm GeV}$ (from bottom to top). 
Also shown is the energy loss rate for neutrino production, 
$Q_{e^- e^+ \rightarrow \nu \bar{\nu}}$, at the same temperature $T_0$.}
\label{fig:Qepm_mrvevr}
\end{figure}
\end{center}

\begin{center}
\begin{figure}[h!]
\includegraphics[width=0.6\textwidth,angle=-90]{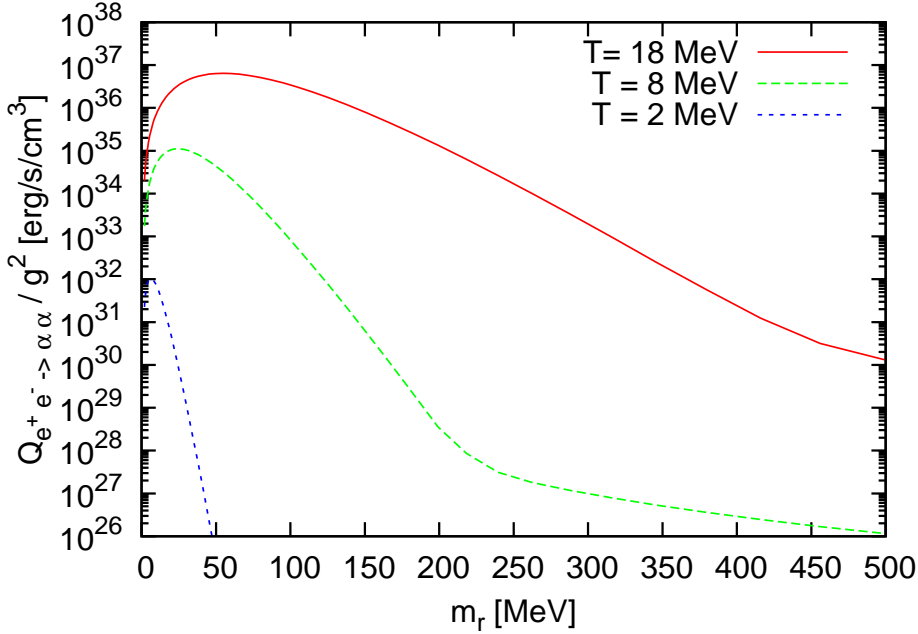}
\caption{Energy loss rate due to Goldstone boson production from 
$e^- e^+ \rightarrow \al \al$ divided by the Goldstone boson coupling $g^2$ vs.
the radial boson mass $m_r$.
The GRB initial fireball temperature is chosen at the fiducial value
$T_0 = 18~{\rm MeV}$, as well as lower values $T_0 = 8$ and $2~{\rm MeV}$,
where the vacuum expectation value of the radial boson is assumed to be 
$\vevr=1~{\rm GeV}$.}
\label{fig:Qepm_mrT}
\end{figure}
\end{center}

The energy loss rate due to Goldstone boson production in the GRB
fireball comoving frame is
\begin{eqnarray}
   Q_{e^+ e^- \rightarrow \al \al} &=& \frac{1}{2!} \int \prod^2_{j=1} 
   \frac{d^3 \vec{q}_j}{\left(2 \pi \right)^3\, 2 \omega_j}
   \int \prod^2_{i=1} \frac{2\, d^3 \vec{p}_i}{\left(2 \pi \right)^3\, 2 E_i}\,
   \frac{1}{4} \sum_{\rm spins} \vert {\cal M}_{e^+ e^- \rightarrow \al \al}\vert^2 
   \nonumber \\
   && \times \left(2 \pi \right)^4 \delta^4 \left(p_1 + p_2 - q_1 - q_2 \right)\, 
   f_{e^-} f_{e^+}\, \left(\omega_1 + \omega_2 \right)\, ,
\end{eqnarray}
where $f_{e^+} \left(\vec{p}_1 \right)$ and $f_{e^-} \left(\vec{p}_2 \right)$ are the 
electron and positron distribution function, respectively, as given below 
Eq.~(\ref{eq:fepm}).
The energy of the two Goldstone bosons in the final state are denoted by
$\omega_1$ and $\omega_2$, while the energy of the positron and electron in the initial
state by $E_1$ and $E_2$.
A symmetry factor of $1/2!$ is included for the two identical particles in the final 
state.
One can perform the $d^3 \vec{q}_1 d^3 \vec{q}_2$ integral analytically analogous to the 
Lenard's Identity~\cite{Lenard:1953zz} for the $e^+ e^- \rightarrow \nu \bar{\nu}$ 
process,
\begin{equation}
   \int \frac{d^3 \vec{q}_1}{\omega_1} \frac{d^3 \vec{q}_2}{\omega_2}\,
   \delta^4 \left(p_1 + p_2 - q_1 - q_2 \right) = 2 \pi\, .
\end{equation}
We use the Maxwell-Boltzmann statistics for the electron and positron distribution 
function, and make a change of integration variables from $E_1$, $E_2$, and 
$\cos \theta$, to
$E_+ \equiv E_1 + E_2$, $E_- \equiv E_1 - E_2$, and $s = 2 m^2_e + 2 E_1 E_2 - 
2 |\vec{p}_1| |\vec{p}_2| \cos \theta$.
The $d E_-$ integral can be performed easily,
\begin{equation} 
   \int^{E^{\rm max}_-}_{E^{\rm max}_-} d E_- = 2 \sqrt{1 - \frac{4 m^2_e}{s}}
   \sqrt{E^2_+ - s}\, .
\end{equation}
Defining $x \equiv E_+ /\sqrt{s}$, $z \equiv s /T^2$, $z_r \equiv m^2_r / T^2$, 
$z_\Gamma \equiv \Gamma^2_r / T^2$, and $z_0 \equiv 4 m^2_e / T^2$, 
the energy loss rate is reduced to the simple form
\begin{equation}
\label{eq:Q_eealal}
   \hspace{-1cm} Q_{e^+ e^- \rightarrow \al \al} = \frac{T^7}{16 \left(2 \pi \right)^5}
   \frac{g^2\, m^2_e}{m^4_\varphi} \int^\infty_{z_0} d z
   \frac{z^{9/2} \left(1 - \frac{z_0}{z} \right)^{3/2}}{\left(z - z_r \right)^2
   + z_r z_\Gamma} \int^\infty_1 d x\, e^{-\sqrt{z} x}\, x \sqrt{x^2 - 1}\, ,
\end{equation}
which we evalulate numerically using the VEGAS Monte Carlo integration
subroutine~\cite{Lepage:1977sw}.
In the resonance region $z \sim z_r$, we simplify the $dz$ integral by taking limit of 
the Poisson kernel
\begin{equation}
\label{eq:poissonkernel}
   \lim_{\epsilon \rightarrow 0} \frac{1}{\pi} \frac{\epsilon}{a^2 + \epsilon^2} = 
   \delta (a)\, .
\end{equation}
Therefore the $dz dx$ integral part can be approximated by
\begin{equation}
\label{eq:poissonI}
  I (T, m_r, \vevr) \approx \frac{z^{9/2}_r\, 
  \left(1 - \frac{z_0}{z_r} \right)^{3/2}\,
  \pi}{\sqrt{z_r z_\Gamma}}\, \int^\infty_1 d x\, e^{- \sqrt{z_r} x}\, 
  x \sqrt{x^2 - 1}\, , 
\end{equation}
for $z \sim z_r$ and $m_r \Gamma_r \ll T^2$.
The results for $T_0 = 18~{\rm MeV}$ and various $m_r$, $\vevr$ values are shown in 
Fig.~\ref{fig:Qepm_mrvevr}.
In the resonance region,
\begin{equation}
\label{eq:Qwidths}
   Q_{e^+ e^- \rightarrow \al \al} \propto \frac{\Gamma_{r \rightarrow e^+ e^-}\,
   \Gamma_{r \rightarrow \al \al}}{m_r \Gamma_r}\, m^5_r\, 
   \int^\infty_1 d x\, e^{- \frac{m_r}{T} x}\, x \sqrt{x^2 - 1}\, .   
\end{equation}
One sees that for a given $m_r$, the Goldstone boson emissivity is enhanced 
significantly due to the resonance effect as 
$Q_{e^+ e^- \rightarrow \al \al} \propto \vevr^2$,
as long as $\Gamma_{r \rightarrow f \bar{f}} \ll \Gamma_{r \rightarrow \al \al}$.
In Fig.~\ref{fig:Qepm_mrT} we show the Goldstone boson emissivity 
$Q_{e^+ e^- \rightarrow \al \al}$ for other GRB initial fireball temperatures than the 
fiducial value $T_0 = 18~{\rm MeV}$, such as $T_0 = 8$ and $2~{\rm MeV}$.
In the resonance region, the $T$-dependence arises solely from the $dx$ integral.
For very large $m_r$ values away from the resonance region, the Goldstone boson 
emissivity depends very sensitively on the GRB fireball temperature as 
\begin{equation}
   Q_{e^+ e^- \rightarrow \al \al} \propto \left(\frac{g^2 m^2_e}
   {m^4_\varphi m^4_r} \right) T^{11}\, .
\end{equation}

\begin{center}
\begin{figure}[t!]
\includegraphics[width=0.6\textwidth,angle=-90]{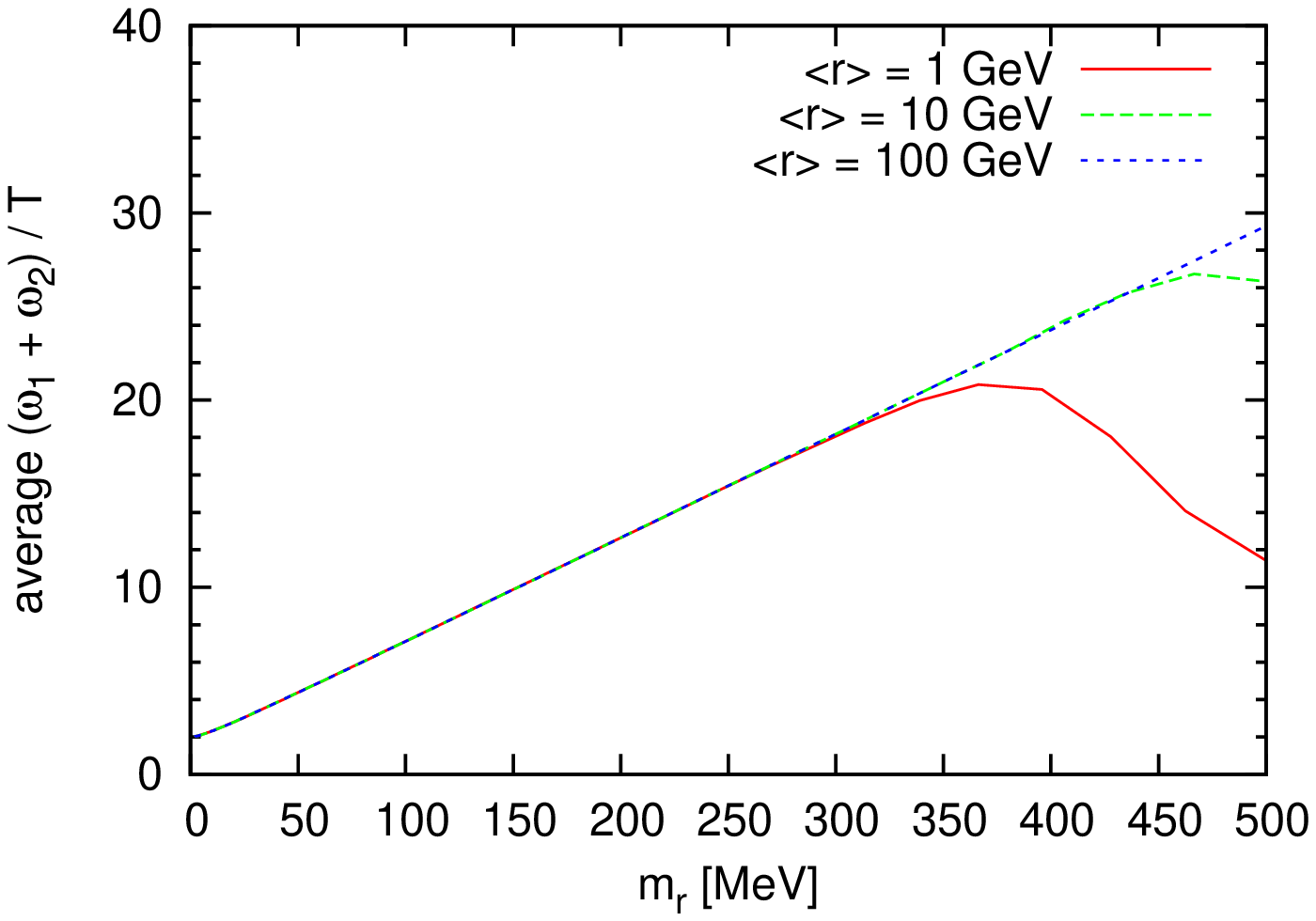}
\caption{Average energy of the emitted Goldstone boson pairs from the process
$e^- e^+ \rightarrow \al \al$ vs. the radial boson mass $m_r$.
The GRB initial fireball temperature is chosen at the fiducial value
$T_0 = 18~{\rm MeV}$,
where the vacuum expectation value of the radial boson is assumed to be $\vevr=1$,
$10$ and $100~{\rm GeV}$.
}
\label{fig:Eepm_mrvevr}
\end{figure}
\end{center}

\begin{center}
\begin{figure}[h!]
\includegraphics[width=0.6\textwidth,angle=-90]{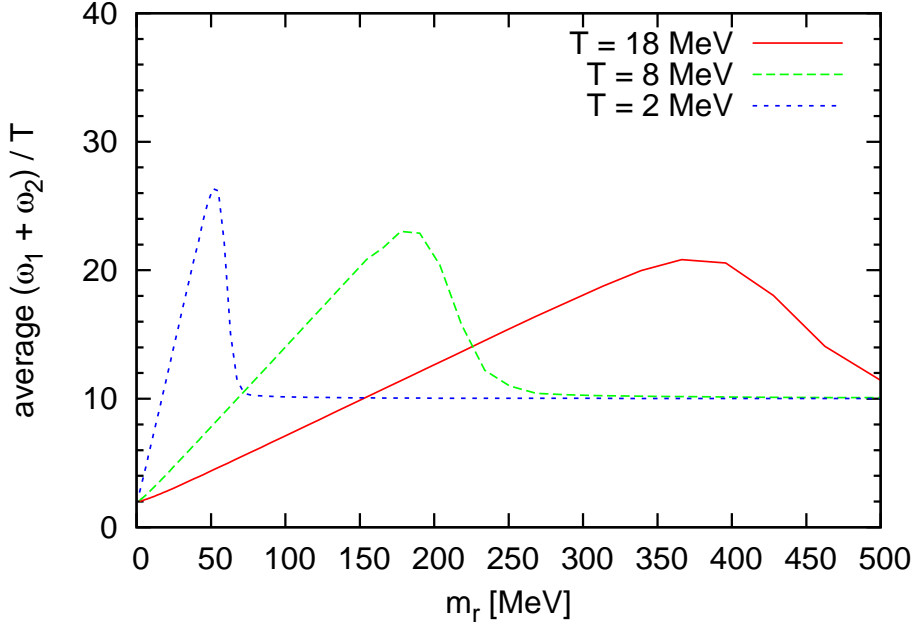}
\caption{Average energy of the emitted Goldstone boson pairs from the process
$e^- e^+ \rightarrow \al \al$ vs. the radial boson mass $m_r$.
The GRB initial fireball temperature is chosen at the fiducial value
$T_0 = 18~{\rm MeV}$, as well as at $T_0 = 8$ and $2~{\rm MeV}$,
where the vacuum expectation value of the radial boson $\vevr$ is assumed to 
be $1~{\rm GeV}$.}
\label{fig:Eepm_mrT}
\end{figure}
\end{center}

As will be presented in Section~\ref{sec:mfp}, the opacity of the GRB fireballs to
the Goldstone bosons depends strongly on the Goldstone boson energy.
The Goldstone boson pairs are emitted with an average energy of
\begin{equation}
   \frac{\bar{\omega}}{T} 
   = \frac{1}{T} \frac{Q_{e^+ e^- \rightarrow \al \al}}{n_{e^-} n_{e^+} 
   \left< \sigma_{e^+ e^- \rightarrow \al \al}\, v_{\rm M} \right>}\, ,
\end{equation}
where $\omega \equiv \omega_1 + \omega_2$, and $v_{\rm M}$ is the M$\o$ller velocity.
The results for $T_0 = 18~{\rm MeV}$ and $\vevr = 1$, $10$, and $100~{\rm GeV}$ are 
shown in Fig.~\ref{fig:Eepm_mrvevr}, while those for $T_0 = 18$, $8$, and $2~{\rm MeV}$ 
and $\vevr = 1~{\rm GeV}$ in Fig.~\ref{fig:Eepm_mrT}.
In the resonance region where the approximation with the Poisson kernel limit in
Eq.~(\ref{eq:poissonI}) is valid, the average energy of the Goldstone boson pairs is 
$\overline{\omega} \propto m_r$.
For large $m_r$ values away from the resonance region, $\omega / T$ approaches a
constant.


\subsection{Photon Scattering}

The amplitude for the photon scattering process
$\gamma (p_1)\, \gamma (p_2) \rightarrow \al (q_1)\, \al (q_2)$ is
\begin{equation}
   \sum_{\rm pol.}|\mathcal{M}_{\gamma \gamma \rightarrow \al \al}|^2 = 
   \left(\frac{\alpha}{4 \pi} \right)^2 \frac{G_F}{\sqrt{2}}\, |F_\gamma|^2\,
   \frac{g^2 \vevh^2}{m^4_\varphi} \frac{s^4}{(s - m^2_r)^2+ m^2_r \Gamma^2_r}\, ,
\end{equation}
where $\alpha$ and $G_F$ are the fine-structure constant an the Fermi constant,
respectively.
The form factor $F_\gamma$ enters through the amplitude for the SM Higgs 
decay to two photons, in this case a function of the centre-of-mass (cm) energy 
$\sqrt{s}$ in the photon collision.
The cm energies attainable at the typical temperature of the initial GRB fireballs  
correspond to the mass of the light (sub-GeV) Higgs boson studied in 
Refs.~\cite{Ellis:1975ap,Leutwyler:1989tn}.
For simplicity, we use a constant value of $|F_\gamma|^2 = 4$ to approximate the result
of Ref.~\cite{Leutwyler:1989tn}.
The energy loss rate is then 
\begin{eqnarray}
   Q_{\gamma \gamma \rightarrow \al \al} &=& \frac{1}{32 \sqrt{2}} 
   \frac{T^9}{\left( 2 \pi \right)^5} \left( \frac{\alpha}{4 \pi} \right)^2 
   G_F \vert F_\gamma \vert^2\, \frac{g^2 \vevh^2}{m^4_\varphi}
   \int^\infty_0 d z \frac{z^{\frac{11}{2}}}{\left(z - z_r \right)^2 + z_r z_\Gamma}
   \nonumber \\
   && \times \int^\infty_1 d x e^{- \sqrt{z} x} x \sqrt{x^2 - 1}\, , 
\end{eqnarray}
where $x$, $z$, $z_r$, and $z_\Gamma$ are defined as in last subsection.
In the resonance region, it can also be expressed in the form of Eq.~(\ref{eq:Qwidths}),
with $\Gamma_{r \rightarrow e^+ e^-}$ replaced by $\Gamma_{r \rightarrow \gamma \gamma}$.
Since the branching ratio for $r \rightarrow \gamma \gamma$ is smaller than $10\%$ of  
that for $r \rightarrow e^+ e^-$ for $m_r \leq 200~{\rm MeV}$, and becomes comparable
only for $m_r \simeq 500~{\rm MeV}$, this process is always subdominant in the parameter 
space we consider in this work.

\subsection{Nuclear Bremsstrahlung Processes}

In the one-pion exchange (OPE) approximation (see e.g. Ref.~\cite{Brinkmann:1988vi}), 
there are four direct and four exchange diagrams, corresponding to the Goldstone boson 
pairs being emitted by any one of the nucleons.
Summing all diagrams and expanding in powers of $\left(T / m_N \right)$, 
the amplitude for the nuclear bremsstrahlung processes
$N (p_1)\, N (p_2) \rightarrow N (p_3)\, N (p_4)\, \al (q_1)\, \al (q_2)$ 
is~\cite{Keung:2013mfa} 
\begin{eqnarray} 
\label{eq:Mbremsstrahlung}
	\sum_{\rm spins} |\mathcal{M}_{N N \rightarrow N N \al \al}|^2 &\approx&
	64 \left(\frac{f_N\, g m_N}{m^2_\varphi} \right)^2 
	\left(\frac{2 m_N f_\pi}{m_\pi} \right)^4 \frac{(q_1 \cdot q_2)^2}
	{(q^2 - m^2_r)^2 + m^2_r \Gamma^2_r} \nonumber \\
	&& \hspace{-4.6cm} \cdot\, \frac{(-2 q^2)^2\, m^2_N}{(2 p \cdot q)^4}\, 
	\Big\{\frac{\vert \vk \vert^4}{(\vert \vk \vert^2 + m^2_\pi)^2} + 
	\frac{\vert \vl \vert^4}{(\vert \vl \vert^2 + m^2_\pi)^2} +
	\frac{\vert \vk \vert^2 \vert \vl \vert^2 - 2 |\vk \cdot \vl|^2}
	{(\vert \vk \vert^2 + m^2_\pi) (\vert \vl \vert^2 + m^2_\pi)} + ... \Big\}\, ,
\end{eqnarray}
with $q \equiv q_1 + q_2$, and $k \equiv p_2 - p_4$ and $l \equiv p_2 - p_3$ are the 
$4$-momenta of the exchanged pion in the direct and the exchange diagrams, respectively.
Here, $\alpha_\pi \equiv \left( 2 m_N f_\pi / m_\pi \right)^2 / 
\left( 4 \pi\right) \approx 15$, with $f_\pi \approx 1$ being the pion-nucleon
``fine-structure" constant.
Goldstone boson pairs can also be emitted from the exchanged pion,
and this contribution is of the same order as Eq.~(\ref{eq:Mbremsstrahlung})
in the $\left(T / m_N \right)$ expansion.
We calculate the energy loss rate in the fireball comoving frame
\begin{eqnarray}
\label{eq:Q_formula}
	Q_{N N \rightarrow N N \al \al} &=& \frac{\mathcal{S}}{2!} 
	\int \frac{d^3 \vec{q_1}}{2 \omega_1\, (2 \pi)^3}
	\frac{d^3 \vec{q_2}}{2 \omega_2\, (2 \pi)^3}\, \int \prod^{4}_{i=1}
	\frac{d^3 \vec{p_j}}{2 E_j\, (2 \pi)^3}\, f_1 f_2 (1-f_3) (1-f_4) \nonumber \\
	&& \hspace{-2cm} \times\, \sum_{\rm spins} 
	|\mathcal{M}_{N N \rightarrow N N \al \al}|^2\, (2 \pi)^4 
	\delta^4 (p_1 + p_2 - p_3 - p_4 - q_1 - q_2)\, (\omega_1 + \omega_2)\, ,
\end{eqnarray}
where $\omega_1, \omega_2$ are the energy of the Goldstone bosons in the final state,
and the distribution functions of the nucleons in the initial and the final 
state are given by
$f_j (\vec{p}_j) = (n_B / 2) (2 \pi / m_N T)^{3/2} e^{- \vert \vec{p}_j
\vert^2 / 2 m_N T}$.
The symmetry factor ${\mathcal S}$ is $\frac{1}{4}$ for $n n $ and $p p$ 
interactions, whereas for $n p$ interactions it is $1$.
We perform the integral over the Goldstone boson momenta first
\begin{equation}
	\hspace{-0.5cm} \int \frac{d^3 \vec{q_1}}{\omega_1}\, 
	\frac{d^3 \vec{q_2}}{\omega_2}\,
	\frac{(q_1 \cdot q_2)^2}{(q^2 - m^2_r)^2 + m^2_r \Gamma^2_r}\, 
	\frac{(2 q^2)^2}{(2 p \cdot q)^4}\, \omega 
	= \frac{2 (2 \pi)^2}{m^4_N} \int^\infty_0 \frac{d \omega}{\omega^3}\, 
	\omega^7\, I_1 (\omega, m_r, \vevr)\, , 	
\end{equation}
where $\omega = \omega_1 + \omega_2$, and the dimensionless integral is
\begin{equation}
\label{eq:NNresonance}
   I_1 (\omega, m_r, \vevr) \equiv \int^1_0 d \tilde{\omega} 
   \int^{+1}_{-1} 
   \frac{d \cos \theta\, \tilom^5\, (1-\tilom)^5\, (1-\cos\theta)^4}{[2 \tilom\, 
   (1-\tilom)\, (1-\cos\theta) - \frac{m^2_r}{\omega^2}]^2 + \frac{m^2_r\, 
   \Gamma^2_r}{\omega^4}}\, .
\end{equation}
Here $\tilom \equiv \omega_1 / \omega$, and $\theta$ is the angle between the two 
emitted Goldstone bosons.
We evaluate Eq.~(\ref{eq:NNresonance}) numerically using the VEGAS subroutine,
and then evaluate the integral in Eq.~(\ref{eq:Q_formula}) over the nucleon momenta 
following Ref.~\cite{Raffelt:1993ix}.
In the non-relativistic limit the nucleon energies are just $E_j = m_N + 
\vert \vec{p}_j \vert^2 / 2 m_N$.
To simplify the nucleon phase space integration, one introduces the centre-of-mass
momenta $\vec{P}$, so that $\vec{p}_{1, 2} = \vec{P} \pm \vec{p}_i$ and 
$\vec{p}_{3, 4} = \vec{P} \pm \vec{p}_f$, as well as 
$z \equiv \vec{p}_i \cdot \vec{p}_f / \vert \vec{p}_i \vert \vert \vec{p}_f \vert$, the 
cosine of the nucleon scattering angle.
The integral over $d^3 \vec{P}$ can be done separately. 
After that one makes a change to dimensionless variables 
$u \equiv \vec{p}^2_i / m_N T$, $v \equiv \vec{p}^2_f / m_N T$, $x \equiv \omega / T$,
and $y \equiv m^2_\pi / m_N T$.
For simplicity we neglect the pion mass $m_\pi$ inside the curly bracket in 
$\sum_{\rm spins} \vert \mathcal{M}_{N N \rightarrow N N \al \al} \vert^2$,
Eq.~(\ref{eq:Mbremsstrahlung}), in comparison with the momentum transfer $\vk$ and 
$\vl$.
The energy loss rate is then
\begin{equation}
	Q_{N N \rightarrow N N \al \al} =  
	\frac{\mathcal{S} \sqrt{\pi}}{(2 \pi)^6}\, (3- \frac{2 \beta}{3})\, 
	I_0\, n^2_B\, \left(\frac{f_N g\, m_N}{m^2_\varphi}\right)^2\, 
	\left(\frac{2 m_N f_\pi}{m_\pi}\right)^4 \cdot \frac{T^{5.5}}{m^{4.5}_N}\, ,
\end{equation}
where we have defined the integral $I_0$ by
\begin{equation}
	I_0 (T, m_r, \vevr) \equiv \int du\, dv\, dx\, x^4\, I_1 (x, m_r, \vevr)\, 
	\sqrt{uv}\, e^{-u}\, \delta (u - v - x)\, ,   
\end{equation}
and the $\beta$ term by
\begin{equation}
   \beta \equiv \frac{3}{I_0} \int du\, dv\, dx\, x^4\, I_1 (x, m_r, \vevr)\, 
	\sqrt{uv}\, e^{-u}\, \delta (u - v - x) \int^{+1}_{-1} \frac{d z}{2}
	\frac{\vert \vk \cdot \vl \vert^2}{\vert \vk \vert^2 \vert \vl \vert^2}\, .
\end{equation}
With the initial comoving baryon number density in the fireball set at the fiducial value
$n_B = 5 \cdot 10^{31}~{\rm cm}^{-3}$,
we find that the energy loss rate due to nuclear bremsstrahlung processes is always
$\sim 10^{-8}$ times that due to electron-positron annihilation process.

\section{Goldstone Boson Mean Free Path in the GRB Fireball} 
\label{sec:mfp}

In this section we estimate the fireball's opacity to the Goldstone bosons. 
The Goldstone boson mean free path in the initial GRB fireball is set by the elastic 
scattering on electrons and positrons $\al + e^\pm \rightarrow \al + e^\pm$, 
as well as on nucleons $\al + N \rightarrow \al + N$.

\subsection{Scattering on Electrons and Positrons}

\begin{center}
\begin{figure}[t!]
\includegraphics[width=0.6\textwidth,angle=-90]{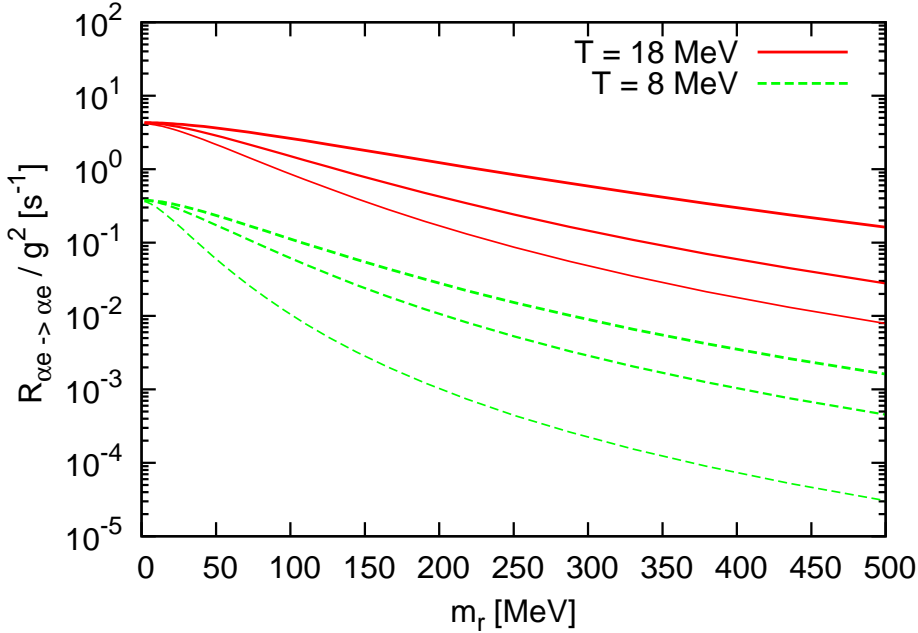}
\caption{The $\al e \rightarrow \al e$ scattering rate divided by the 
coupling $g^2$ versus the radial boson mass $m_r$. 
The initial GRB fireball temperature is assumed to be at the fiducial value 
$T_0 = 18~{\rm MeV}$, where we show the results for a Goldstone boson with incident 
energy $\omega_1 = 540$, $180$, and $90~{\rm MeV}$ (solid lines, from top to bottom.)
The results for $T_0 = 8~{\rm MeV}$ are also displayed, for $\omega_1 = 320$, $160$, and 
$40~{\rm MeV}$ (dashed lines, from top to bottom.)}
\label{fig:Raleale}
\end{figure}
\end{center}

The amplitude for Goldstone boson scattering on electrons and positrons 
$\al (q_1)\, e^\pm (p_1) \rightarrow \al (q_2)\, e^\pm (p_2)$ is
\begin{equation}
   \sum_{\rm spins} \vert {\cal M}_{\al e \rightarrow \al e} \vert^2 = 
   \frac{4 g^2 m^2_e}{m^4_\varphi}
   \frac{\left(q_1 \cdot q_2 \right)^2\, \left[(p_1 \cdot p_2) + m^2_e \right]}
   {\left(t - m^2_r \right)^2}\, ,
\end{equation}
where $t = (q_2 - q_1)^2 = (p_1 - p_2)^2$. 
We follow Ref.~\cite{Tubbs:1975jx} to calculate the reaction rate 
\begin{eqnarray}
   R_{\al e \rightarrow \al e} &=& n_e\, \sigma_{\al e \rightarrow \al e}\, 
   \vM 
   = \int \frac{2 d^3\vec{p}_1}{(2 \pi)^3} f_e (\vec{p}_1)\, 
   \frac{1}{2 \omega_1\, 2 E_1}\, 
   \int \frac{d^3\vec{q}_2}{(2 \pi)^3\, 2 \omega_2} \nonumber \\
   && \hspace{-2.5cm} \times \int \frac{d^3\vec{p}_2}{(2 \pi)^3\, 2 E_2}\,
   \left[1 - f_e (\vec{p}_2) \right]\,    
   \frac{1}{2} \sum_{\rm spins} |\mathcal{M}_{\al e \rightarrow \al e}|^2\, 
   (2 \pi)^4 \delta^4 (p_1 + q_1 - p_2 - q_2)\, .
\end{eqnarray}
Using the polar angle $\cos \theta \equiv \vec{p}_1 \cdot \vec{q}_1 / |\vec{p}_1| 
|\vec{q}_1|$ and the azimuthal angel $\phi^\prime$ which is measured from the 
$(\vec{p}_1, \vec{q}_1)$-plane, the 9-dimensional integral can be simplified to
\begin{eqnarray}   
\label{eq:Raleale}
   R_{\al e \rightarrow \al e} &=& \frac{1}{(2 \pi)^3}\, \frac{m^4_e}{4 \omega_1} 
   \frac{g^2 m^2_e}{m^4_\varphi} 
   \int^\infty_1 d \epsilon_1 f_e (\epsilon_1) \sqrt{\epsilon^2_1 - 1} 
   \int^{+1}_{-1} \frac{d \cos \theta}{\lambda (\epsilon_1, u_1, \cos \theta)} 
   \nonumber \\
   && \times \int^{\epsilon^{\rm max}_2}_{\epsilon^{\rm min}_2}  d \epsilon_2
   \left[1 - f_e (\epsilon_2) \right] \int^{2 \pi}_0 \frac{d \phi^\prime}{2 \pi}
   F_3\, ,
\end{eqnarray}
with the dimensionless variables 
$\epsilon_1 \equiv E_1 / m_e$, $\epsilon_2 \equiv E_2 / m_e$, 
and $u_1 \equiv \omega_1 / m_e$. 
The functions in the above equation are defined as
\begin{equation}
   \lambda (\epsilon_1, u_1, \cos \theta) \equiv \frac{|\vec{p}_1 + \vec{q}_1|}{m_e} 
   = \sqrt{\epsilon^2_1 - 1 + u^2_1 + 2 u_1 \left(\epsilon^2_1 - 1 \right)^{1/2} 
   	\cos \theta}\, ,
\end{equation}
and
\begin{equation}
   F_3 \equiv \frac{\left[q_1 \cdot \left(p_1 + q_1 - p_2 \right) \right]^3
   	+ 2 m^2_e \left[q_1 \cdot \left(p_1 + q_1 - p_2 \right) \right]^2}
   {\left[ 2 q_1 \cdot \left(p_1 + q_1 - p_2 \right) + m^2_r \right]^2\, m^2_e}\, ,
\end{equation}
respectively, and the limits for the $d \epsilon_2$ integration are determined to be
\begin{equation}
   \epsilon^{\rm max , \, \min}_2 = \frac{1}{2}\, \left[\epsilon_1 + u_1 \pm
   \lambda (\epsilon_1, u_1, \cos \theta) + \frac{1}
   {\epsilon_1 + u_1 \pm \lambda (\epsilon_1, u_1, \cos \theta)} \right]\, .
\end{equation}
To evaluate $q_1 \cdot p_2$, we need to know the angle
\begin{equation}
   \cos \theta_{q_1 p_2} \equiv \cos \theta^\prime \cos \Delta_2 - 
   \sin \theta^\prime \sin \Delta_2 \cos \phi^\prime\, ,	 
\end{equation}	
where
\begin{equation}
   \cos \Delta_1 = \frac{\sqrt{\epsilon^2_1 - 1} + u_1 \cos \theta}{\lambda}\, ,
   \hspace{0.4cm}
   \cos \Delta_2 = \frac{u_1 + \sqrt{\epsilon^2_1 - 1} \cos \theta}{\lambda}\, ,
\end{equation}
with $\Delta_1 + \Delta_2 = \theta$.
We evaluate Eq.~(\ref{eq:Raleale}) numerically using the VEGAS subroutine.
In Fig.~\ref{fig:Raleale} we plot the $\al e \rightarrow \al e$ scattering rate divided 
by the Goldstone boson coupling $g^2$, for an incident Goldstone boson energy of 
$\omega_1 = 540$, $180$, and $90~{\rm MeV}$, assuming the fiducial initial GRB 
fireball temperature $T_0 = 18~{\rm MeV}$.
The rates for $T_0 = 8~{\rm MeV}$ are also displayed, for Goldstone boson incident energy
$\omega_1 = 320$, $160$, and $40~{\rm MeV}$.
We find that for all Goldstone boson energies attainable in the GRB initial fireballs 
and all $m_r$ values, $R_{\al e \rightarrow \al e} \lesssim 4 g^2~{\rm s}^{-1}$.

\subsection{Scattering on Nucleons}

\begin{center}
\begin{figure}[t!]
\includegraphics[width=0.6\textwidth,angle=-90]{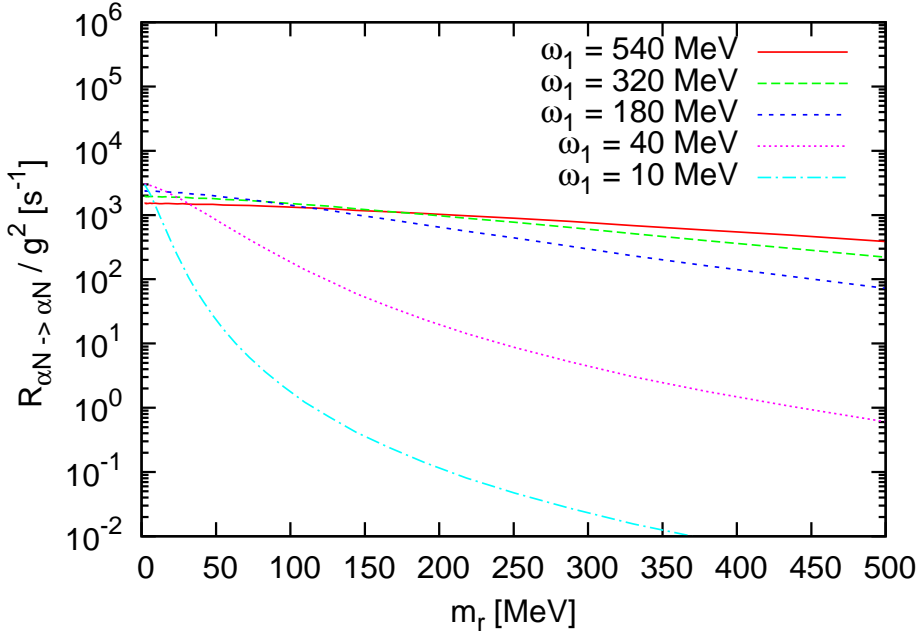}
\caption{Results for the $\al N \rightarrow \al N$ scattering rate divided by the 
coupling $g^2$ versus the radial field mass $m_r$, 
for incident Goldstone boson energy $\omega_1 = 540$, 
$320$, $180$, $40$, and $10~{\rm MeV}$ (from top to bottom).
The baryon density is assumed to be $n_B = 5 \cdot 10^{31}~{\rm cm}^{-3}$.}
\label{fig:RalNalN}
\end{figure}
\end{center}

The interaction rate for $\al (q_1) N (p_1) \rightarrow \al (q_2) N (p_1)$ can be 
calculated similarly as $R_{\al e \rightarrow \al e}$ in Eq.~(\ref{eq:Raleale}) by
replacing $m_e$ with $m_N$ and using the non-relativistic Maxwell-Boltzmann distribution
$f_N \left( \vec{p} \right) = \left( n_B / 2 \right)\, 
\left( 2 \pi / m_N T \right)^{3/2}\, e^{- \vec{p}^2 / 2 m_N T}$ for the nucleons.
The amplitude squared is
\begin{equation}
	\sigma_{\al N \rightarrow \al N} = \frac{4 f^2_N g^2\, m^2_N}{m^4_\varphi}
	\frac{(q_1 \cdot q_2)^2\, \left[\left(p_1 \cdot p_2 \right) + m^2_N \right]}
	{(t - m^2_r)^2}\, ,
\end{equation}
where $g_N$ is the effective coupling of the Goldstone bosons to nucleons.
For low incident Goldstone boson energies $\omega_1 \ll m_N$, the nuclear recoil effects
can be neglected, and so the interaction rate can also be easily estimated by
\begin{eqnarray}
   R_{\al N \rightarrow \al N} &=& n_B\, \sigma_{\al N \rightarrow \al N}\, \vM
   \nonumber \\
   &=& n_B \frac{\omega^4_1}{16 \pi} \frac{f^2_N g^2}{m^4_\varphi} \int^{+1}_{-1} 
   d \cos \theta \frac{\omega^2_1 \left(1 - \cos \theta \right)^3 + 2 m^2_N
   \left(1 - \cos \theta \right)^2}{\left[2 \omega^2_1 \left(1 - \cos \theta \right)
   + m^2_r \right]^2}\, .
\end{eqnarray}
We found that the results from this method agree with those from the full calculation 
within $10\%$ for $\omega_1 \lesssim 40~{\rm MeV}$.
The results are shown in Fig.~\ref{fig:RalNalN}, where we assume the baryon number 
density in the GRB fireball is $n_B = 5 \cdot 10^{31}~{\rm cm}^{-3}$.
Although the baryon number density is four orders of magnitude smaller than that of the
electrons and positrons, due to the large nucleon mass $m_N$, this channel
dominates over the scattering on electrons and positrons.  
The figure indicates that there is an upper bound on the scattering rate, 
$R_{\al N \rightarrow \al N} \lesssim 4 \cdot 10^4
\left(f_N g \right)^2~{\rm s}^{-1}$.

With $f_N \sim 0.3$, the Goldstone boson mean free path in the initial GRB fireball 
is then
\begin{equation}
\label{eq:mfp}
  \lambda_{\al} =\left[n_e \sigma_{\al e \rightarrow \al e} + 
  n_B \sigma_{\al N \rightarrow \al N} \right]^{-1} 
  \simeq \left(R_{\al N \rightarrow \al N} \right)^{-1}
  \gtrsim \frac{9.56 \cdot 10^{6}}{g^2}~{\rm cm}\, ,
\end{equation}
for all $m_r$ and $\omega_1$ values.
Taking into account the current collider constraint of $|g| < 0.011$, we find that
\begin{equation}
   \lambda_{\al} \gtrsim 7.9 \times 10^{10}~{\rm cm} \gg R_0\, ,
\end{equation}
for all $m_r$ values.
We conclude that the Goldstone bosons produced in the initial fireball of GRBs 
cannot be trapped therein, i.e. the GRB initial fireballs are transparent to the 
Goldstone bosons.
The consequence will be discussed in the next Section.

\section{Hydrodynamics of GRB fireballs in the Presence of Goldstone Boson 
Production}
\label{sec:hydro}

We apply the relativistic hydrodynamics for describing quark-gluon plasma
anticipated at the LHC or the Relativistic Heavy Ion Collider 
(RHIC)~\cite{RHIC} (see e.g. Ref.~\cite{Hwa:1990xg} for review articles on this topic) to 
study the GRB fireballs.

\subsection{Hydrodynamics of GRB Fireballs with Dissipation}

The evolution of GRB fireballs is governed by the equations for the conservation of 
(baryon) particle number and for the conservation of energy and momentum,  
\begin{equation}
   \partial_\mu N^\mu = 0\, , \hspace{0.4cm}
   \partial_\mu T^{\mu \nu} = j^\nu\, ,
\end{equation}
respectively.
Here $j^\nu$ represents an effective source term, with a negative 
(positive) $j^0$ term denoting an energy sink (source).
The baryon number flux is $N^\mu = n_B u^\mu$.
For each particle species $i$ in the fluid, one expands its phase space 
distribution function around the equilibrium value, as $f = f_0 + \delta f$.
The deviation from the equilibrium value is related to a characteristic 
relaxation time.
With such a correction, the stress-energy tensor is then 
modified to (see e.g. Ref.~\cite{Betz:2010cx})
\begin{equation}
   T^{\mu \nu} = \left(\epsilon_0 + \delta \epsilon \right) u^\mu u^\nu + 
   (p_0 + \pi_b) \Delta^{\mu \nu} + \pi^{\mu \nu}\, ,
\end{equation}
with $u^\mu$ the four-velocity, and 
$\Delta^{\mu \nu} = g^{\mu \nu} + u^\mu u^\nu$ the project tensor to the 
subspace orthogonal to the fluid velocity.
Here we choose the signature of the metric to be $(-, +, +, +)$, and the 
fluid four-velocity $u^\mu$ is specified using the definition by Landau and
Lifshitz. 
Following this definition, the tensor equation $\pi_{\mu \nu} u^\nu = 0$ must
be satisfied.

The shear tensor and the bulk viscosity pressure in the lowest order of the 
velocity gradients are of the form
\begin{equation} 
    \hspace{-1.4cm} \pi^{\mu \nu} = - 2 \eta \left(\frac{1}{2} 
    \left(\Delta^{\mu \alpha} \partial_\alpha u^\nu + 
    \Delta^{\nu \alpha} \partial_\alpha u^\mu \right) - 
    \frac{1}{3} \Delta^{\mu \nu} \partial_\alpha u^\alpha \right)\, ,
    \hspace{0.6cm} \pi_b = - \zeta \partial_\mu u^\mu\, ,
\end{equation}
respectively,
with $\eta$ and $\zeta$ denoting the shear and the bulk viscosity coefficient. 
However, as mentioned in Ref.~\cite{Denicol:2012es}, to avoid the acausality 
problems, the dissipative fields should be regarded as independent dynamical 
variables.
The shear viscosity can be estimated using the Green-Kubo 
relation~\cite{Kubo:1957mj} (see also, e.g. Ref.~\cite{Plumari:2012ep} for a 
recent numerical study.)
From kinetic theory, the shear viscosity coefficient is 
(see e.g. Ref.~\cite{Danielewicz:1984ww})
\begin{equation}
\label{eq:shear}
   \eta \approx \frac{1}{3}\, \sum_j n_j \left< p \right>_j \lambda_j\, ,
\end{equation}
i.e. it is determined by particle species $j$ in the fluid with number density 
$n_j$ transporting an average momentum $\left<p \right>_j$ over a momentum
transport mean free path $\lambda_j$.

To solve the equation for the conservation of energy and momentum, one 
projects it in the direction of the fluid velocity and that orthogonal to
the fluid velocity, obtaining (see e.g. Ref.~\cite{Floerchinger:2014yqa})
\begin{equation}
\label{eq:hydro_para}
   u^\mu \partial_\mu \epsilon + \left(\epsilon + p_0 \right) \partial_\mu u^\mu
   - u_\nu \partial_\mu \pi^{\mu \nu} + \pi_b \partial_\mu u^\mu
   = - u_\nu j^\nu\, .
\end{equation}
and
\begin{equation}
\label{eq:hydro_orth}
   \left(\epsilon + p_0 + \pi_b \right) u^\mu \partial_\mu u^\alpha
   + \Delta^{\alpha \beta} \partial_\beta (p_0 + \pi_b) +
   \Delta^\alpha_{~\nu} \partial_\mu \pi^{\mu \nu} = 
   \Delta^\alpha_{~\nu} j^\nu\, ,
\end{equation}
with $\epsilon = \epsilon_0 + \delta \epsilon$.
These conservation equations need to be supplemented with an equation of state
for the fireball plasma.
When the GRB fireball expansion reaches the coasting phase, i.e. the Lorentz
factor $\Gamma$ is constant, one can transform to the Milne coordinates
\begin{equation}
    \tau \equiv \sqrt{t^2 - R^2}\, , \hspace{0.5cm}
    \chi \equiv \tanh^{-1} \left( R / t \right)\, ,
\end{equation}
as in e.g. Refs.~\cite{Charng:2002ak,Sanches:2015vra}.

The effects of the dissipation fields are to transfer the kinetic energy
into heat, while the energy source (sink) increase (decrease) the total energy.
In the initial fireball of GRBs, we can assume that all particle species -
the electrons and positrons, photons, as well as the protons and neutrons -
are strongly coupled and thus are all in thermal equilibrium.
Now consider the case that from their interactions some exotic particle species 
are copiously created. 
If they are not fully thermalised, they lead to a slower expansion of the 
fireball.

However, in the last section we found that the Goldstone boson mean free path
$\lambda_\al$ exceeds the size of the initial fireball $R_0$
(cf. Eq.~(\ref{eq:mfp})).
The Goldstone bosons produced therein are not trapped and therefore are not
thermalised at all.
In this case Eq.~(\ref{eq:shear}) is not applicable, since its validity requires
$\lambda_\al \ll R_0$.
The effects of the Goldstone bosons can still be estimated by transforming
to the fireball comoving frame.
Following the definition by Landau and Lifshitz, in this frame 
the terms involving $\pi^{\mu \nu}$ or $\pi_b$ completely vanish.

\subsection{The GRB Fireball Energy Loss Criterion}

In the fireball comoving frame, we demand that the Goldstone bosons transport 
away an amount of energy comparable to the initial fireball radiation energy 
before their emissivity decreases significantly with the temperature.
In the GRB fireball comoving frame where the four-velocity is
$u^\nu = (1, 0, 0, 0)$ in spherical coordinates $(t^\prime, R^\prime,
\theta^\prime, \phi^\prime)$, the hydrodynamic equations 
Eq.~(\ref{eq:hydro_para}) and (\ref{eq:hydro_orth}) are simply
\begin{equation}
\label{eq:hydrodynamic_comoving}
   \frac{\partial n_B}{\partial t^\prime} = 0\, , \hspace{0.7cm}
   \frac{\partial \epsilon}{\partial t^\prime} = j^0 = Q\, , \hspace{0.7cm}
   \frac{\partial p}{\partial R^\prime} = 0\, ,
\end{equation}
where the coordinates in the comoving frame and in the observer frame are 
related by the Lorentz factor, i.e. 
$t^\prime = t / \Gamma$ and $R^\prime = \Gamma R$.
Here the baryon number density $n_B$, the energy density $\epsilon$ and the 
the pressure $p$, as well as the energy loss or creation per unit volume per 
unit time $Q$, are all comoving quantities. 
The Goldstone bosons are emitted isotropically in the fireball comoving frame,
so the net momentum flux herein is $j^1 = j^2 = j^3 = 0$.
One can regard the Goldstone bosons as an energy sink.
Using the equation for energy conservation in 
Eq.~(\ref{eq:hydrodynamic_comoving}), we can derive a constraint on the 
Goldstone boson emissivity in the GRB initial fireball as
\begin{equation} 
\label{eq:Qcriterion}
   | \Delta \epsilon | = | - Q_{e^+ e^- \rightarrow \al \al}\, \Delta t^\prime |
   \approx Q_{e^+ e^- \rightarrow \al \al}\, \frac{1}{\Gamma_0} 
   \frac{\Delta R_0}{\beta_0} \gtrsim \frac{\mathcal{E}}{\Gamma_0 V_0}\, .
\end{equation}
Choosing $\Delta R_0 \sim R_0$, this criterion is equivalent to the comparison of the 
cooling timescale $t_c$ with the fireball expansion timescale $t_e$ in 
Ref.~\cite{Koers:2005ya}
\begin{equation}
\label{eq:chiKoers}
   \chi \equiv \frac{t_c}{t_e} \approx 
   \frac{\mathcal{E} / \left(Q_{e^+ e^- \rightarrow \al \al} V_0 \right)}
   {R_0 / \beta_0} \lesssim 1\, .
\end{equation}

\begin{center}
\begin{figure}[t!]
\includegraphics[width=0.6\textwidth,angle=-90]{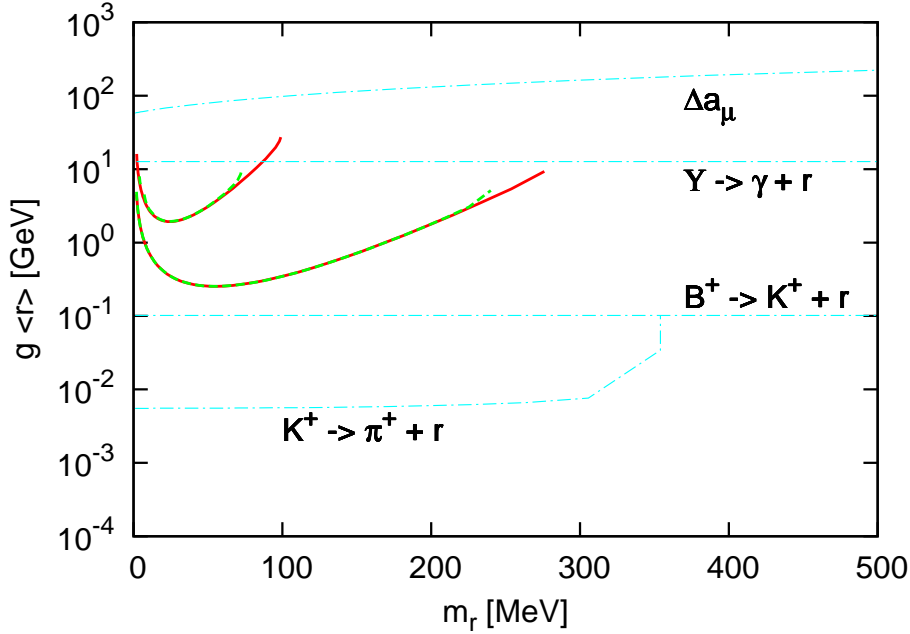}
\caption{Upper limits on $g \vevr$, the product of the coupling with the vev of the 
radial field $r$, versus its mass $m_r$, from the energy loss rate in the GRB initial 
fireball, Eq.~(\ref{eq:Qcriterion}).
The initial fireball temperature is chosen at the fiducial value 
$T_0 = 18~{\rm MeV}$ (lower solid line), as well as lower value $T_0 = 8~{\rm MeV}$ 
(upper solid line), the initial radius at $R_0 = 10^{6.5}~{\rm cm}$, and the 
fireball energy at $\mathcal{E} = 10^{52}~{\rm erg}$.
Here we assume that the Goldstone boson coupling saturates the collider bound of 
$g = 0.011$. 
For a smaller coupling $g = 0.0015$ which might be probed by future collider experiments,
the upper limits are less stringent for $m_r > 240~{\rm MeV}$ if 
$T_0 = 18~{\rm MeV}$ (lower dashed line), or for $m_r > 70~{\rm MeV}$ if
$T_0 = 8~{\rm MeV}$ (upper dashed line).
Also shown are the upper limits from muon anomalous magnetic moment $\Delta a_\mu$, 
radiative Upsilon decays $\Upsilon (nS) \rightarrow \gamma + r$, 
$B^+$ invisible decay $B^+ \rightarrow K^+ r$, as well as $K^+$ invisible decay 
$K^+ \rightarrow \pi^+ r$ (dash-dotted lines, from top to bottom.)}
\label{fig:GRBmeson_gcoupvevr}
\end{figure}
\end{center}

In Fig.~\ref{fig:GRBmeson_gcoupvevr} we plot the upper limits on $g \vevr$, 
the Goldstone boson coupling times the vev of the $r$ field, versus
its mass, $m_r$, obtained by using the criterion in Eq.~(\ref{eq:Qcriterion}).
The GRB initial fireball temperature, radius, and energy, are chosen
at the fiducial value $T_0 = 18~{\rm MeV}$, $R_0 = 10^{6.5}~{\rm cm}$, and
$\mathcal{E} = 10^{52}~{\rm erg}$, as well as a lower initial temperature  
$T_0 = 8~{\rm MeV}$.
If the temperature of the GRB initial fireball is as low as $T_0 = 2~{\rm MeV}$, no
constraint on the parameters of Weinberg's Higgs portal model can be obtained.

In fact, the GRB bounds on $g \vevr$ have a slight dependence on the Goldstone boson 
coupling $g$, which becomes visible when $\Gamma_{r \rightarrow f \bar{f}}$ 
is no longer negligible compared to $\Gamma_{r \rightarrow \al \al}$.
Here we consider $g = 0.011$ saturating current collider bounds, as well as  
$g = 0.0015$ which might be probed by future collider experiments.
For the latter case, the upper limits are less stringent for $m_r > 240~{\rm MeV}$
if $T_0 = 18~{\rm MeV}$, or for $m_r > 70~{\rm MeV}$ if $T_0 = 8~{\rm MeV}$.

An inspection of Fig.~\ref{fig:GRBmeson_gcoupvevr} indicates that in the mass range
$m_r / T_0 \lesssim 10-15$, 
the GRB bounds are indeed competitive to current laboratory constraints 
reviewed in Section~\ref{sec:labconstraints}.
They are more stringent than the constraints from muon anomalous magnetic moment and 
radiative upsilon decays, while weaker than those from the $B^+$ and $K^+$ meson 
invisible decays by $1$-$3$ orders of magnitude.

\section{Summary}
\label{sec:summary}

We aimed to study the effects of the Goldstone bosons in Weinberg's Higgs portal model
on the initial fireballs of gamma-ray bursts.
We first calculated the energy loss rates therein due to Goldstone boson production in 
different channels, including electron-positron annihilation, photon scattering, and
nuclear bremsstrahlung processes.
We found that resonance effects significantly enhance the energy loss rate 
for the electron-positron annihilation process, even for the mass of the radial field 
$r$ approaching $30-40$ times the initial GRB fireball temperature.
On the other hand, in the calculation of the Goldstone boson mean free path, 
there is no such effect present in the processes of Goldstone boson scattering 
on nucleons and electrons or positrons. 
Interestingly, we found that although nuclear bremsstrahlung processes are of no 
importance in Goldstone boson production, the scattering on nucleons dominates over 
scattering on electrons and positrons by four orders of magnitude in setting the 
Goldstone boson mean free path in the GRB fireballs. 
However, for all Goldstone boson energies attainable in the GRB initial fireballs and 
all $m_r$ values, the Goldstone boson mean free path always exceeds the initial fireball
radius.
Thus the Goldstone bosons do not couple to the GRB fireball plasma.
The initial GRB fireballs are transparent to the Goldstone bosons, so that they freely
transport the fireball energy away and act as an energy sink.

We obtained constraints on $g \vevr$, the Goldstone boson coupling times the vacuum 
expectation value of the $r$ field, 
by using the energy loss rate criterion derived from the hydrodynamic equations 
in the GRB fireball comoving frame.
Assuming generic values for the GRB initial fireball temperature, radius and energy,
we found that in the mass range of $m_r / T_0 \lesssim 10-15$, the GRB bounds are indeed
competitive to current laboratory constraints.
They are more stringent than the constraints from muon anomalous magnetic moment and 
radiative upsilon decays, while weaker than those from the $B^+$ and $K^+$ meson 
invisible decays by $1-3$ orders of magnitude.

\section*{Acknowledgements}

This work was supported in part by the Ministry of Science and Technology,
Taiwan, ROC under the Grant No. 104-2112-M-001-039-MY3.

%\section*{References}
%\bibliographystyle{plain}
%\bibliography{mybib}

\end{document}